\documentclass[11pt, oneside]{article}   	% use "amsart" instead of 
\usepackage{geometry}                		% See geometry.pdf to learn the
\usepackage{graphicx}						% Use pdf, png, jpg, or eps with
\usepackage{amssymb}
\usepackage{amsmath,amssymb}
\usepackage{amsthm}
\usepackage{mathtools}
\newcommand\numberthis{\addtocounter{equation}{1} \tag{\theequation}}

\usepackage{times}

\usepackage{xcolor}
\usepackage{fontawesome}
\usepackage[colorlinks]{hyperref}
\usepackage[nameinlink, capitalize]{cleveref}
\usepackage{booktabs} % for tables
\usepackage{enumitem}
\usepackage{lineno}
\usepackage{setspace}
\usepackage{siunitx}
\usepackage{textcomp}
\usepackage[open,atend,openlevel=1,numbered]{bookmark}
\usepackage{xspace}
\usepackage[nonumberlist,nogroupskip,acronym]{glossaries}
\usepackage{caption}

\newcommand{\corrAuthor}{\textsuperscript{\faEnvelopeO}}
\newcommand{\algoname}{M\textsuperscript{2}NuFFT\xspace}

\newglossarystyle{abbtab}{%
	\renewenvironment{theglossary}%
	{\begin{longtable}{p{0.15\linewidth} p{0.6\linewidth}}
			\hline
			\hline
			\textbf{Abbreviation} & \textbf{Description} \\
			\hline\endhead}%
			{\hline\end{longtable}}%
}

\newacronym{bg}{BG}{Bronez GPSS}
\newacronym{dft}{DFT}{Discrete Fourier Transform}
\newacronym{dpss}{DPSS}{Discrete Prolate Spheroidal Sequence}
\newacronym{gep}{GEP}{Generalized Eigenvalue Problem}
\newacronym{gpss}{GPSS}{Generalized Prolate Spheroidal Sequence}
\newacronym{m2nufft}{\algoname}{Multiband-Multitaper Nonuniform Fast Fourier
	Transform}
\newacronym{mdss}{MDSS}{Missing-Data Slepian Sequence}
\newacronym{psd}{PSD}{Power Spectral Density}
\newacronym{mtls}{MTLS}{Multitaper Lomb-Scargle periodogram}
\newacronym{mwmv}{MWMV}{Multiple Window Minimum Variance}
\newacronym{nufft}{NuFFT}{Nonuniform Fast Fourier Transform}

\newglossarystyle{symbtab}{%
	{\begin{longtable}{p{0.15\linewidth} p{0.6\linewidth}}
			\hline
			\hline
			\textbf{Symbol} & \textbf{Description} \\
			\hline\endhead}%
			{\hline\end{longtable}}%
}

\newglossary[slg]{symbols}{sym}{sbl}{}

\newglossaryentry{A}{
	type=symbols,
	name={\ensuremath{\mathcal{A}}},
	description={Generic analysis band}
}

\newglossaryentry{A0}{
	type=symbols,
	name={\ensuremath{\mathcal{A}_0}},
	description={Analysis band centered at $f_{c_0}$, or nominal band}
}

\newglossaryentry{Ai}{
	type=symbols,
	name={\ensuremath{\mathcal{A}_i}},
	description={Analysis band centered at $f_{c_i}$, for $i=1,2,\ldots,I-1$}
}

\newglossaryentry{B}{
	type=symbols,
	name={\ensuremath{\mathcal{B}}},
	description={Entire signal band of signal $\mathbf{x}$}
}

\newglossaryentry{beta}{
	type=symbols,
	name={\ensuremath{\beta}},
	description={Sampling interval of missing data sampling scheme}
}

\newglossaryentry{dpsskn}{
	type=symbols,
	name={\ensuremath{v_n^{(k)}(N,f_w)}},
	description={$n$-th element of the vector with portion of DPSS, order $k$}
}

\newglossaryentry{E}{
	type=symbols,
	name={\ensuremath{\mathcal{E}_i}},
	description={Suboptimality measure at analysis band $\mathcal{A}_i$}
}

\newglossaryentry{Ei}{
	type=symbols,
	name={\ensuremath{E_i}},
	description={Frequency shift operator}
}

\newglossaryentry{eigencoeff}{
	type=symbols,
	name={\ensuremath{J_k(f)}},
	description={$k$-th eigenvalue coefficient of uniformly sampled signal}
}

\newglossaryentry{eigencoeffAi}{
	type=symbols,
	name={\ensuremath{J_k(\mathcal{A}_i)}},
	description={$k$-th eigenvalue coefficient of analysis band $\mathcal{A}_i$}
}

\newglossaryentry{eigenspectrum}{
	type=symbols,
	name={\ensuremath{\left|J_k(f)\right|^2}},
	description={$k$-th eigenspectrum of uniformly sampled signal}
}

\newglossaryentry{eigenspectrumAi}{
	type=symbols,
	name={\ensuremath{\left|J_k(\mathcal{A}_i)\right|^2}},
	description={$k$-th eigenspectrum of analysis band $\mathcal{A}_i$}
}

\newglossaryentry{fc}{
	type=symbols,
	name={\ensuremath{f_c}},
	description={Center frequency of generic analysis band $\mathcal{A}$}
}

\newglossaryentry{fci}{
	type=symbols,
	name={\ensuremath{f_{c_i}}},
	description={Center frequency of analysis band $\mathcal{A}_i$, for $i=0,1,\ldots,I-1$}		}

\newglossaryentry{fmax}{
	type=symbols,
	name={\ensuremath{f_{\max}}},
	description={Maximum frequency of signal band $\mathcal{B}$},
}

\newglossaryentry{fs}{
	type=symbols,
	name={\ensuremath{f_s}},
	description={Sampling frequency}
}

\newglossaryentry{gmatA0}{
	type=symbols,
	name={\ensuremath{\mathbf{R}(\mathcal{A}_0)}},
	description={GPSS matrix for analysis band $\mathcal{A}_0$}
}

\newglossaryentry{gmatAi}{
	type=symbols,
	name={\ensuremath{\mathbf{R}(\mathcal{A}_i)}},
	description={GPSS matrix for analysis band $\mathcal{A}_i$, for $i=0,1,\ldots,I-1$}
}

\newglossaryentry{gmatB}{
	type=symbols,
	name={\ensuremath{\mathbf{R}(\mathcal{B})}},
	description={GPSS matrix for signal band $\mathcal{B}$}
}

\newglossaryentry{gmatAnm}{
	type=symbols,
	name={\ensuremath{\mathbf{R}(\mathcal{A}| n, m)}},
	description={Element $(n,m)$ of GPSS matrix $\mathbf{R}(\mathcal{A})$}
}

\newglossaryentry{gmatBnm}{
	type=symbols,
	name={\ensuremath{\mathbf{R}(\mathcal{B}| n, m)}},
	description={Element $(n,m)$ of GPSS matrix $\mathbf{R}(\mathcal{B})$}
}

\newglossaryentry{halfbw}{
	type=symbols,
	name={\ensuremath{f_w}},
	description={Bandwidth of an analysis band $\mathcal{A}$, whose frequency resolution is $2 f_w$.}
}

\newglossaryentry{hatE}{
	type=symbols,
	name={\ensuremath{\hat{\mathcal{E}}_i}},
	description={Estimate of suboptimality measure $\mathcal{E}_i$}
}

\newglossaryentry{hatLki}{
	type=symbols,
	name={\ensuremath{\hat{L}_k^i}},
	description={Estimate of taper suboptimality measure $L_k^i$}
}

\newglossaryentry{idmat}{
	type=symbols,
	name={\ensuremath{\mathbb{I}}},
	description={Identity matrix}
}

\newglossaryentry{intspectrum}{
	type=symbols,
	name={\ensuremath{\hat{P}(\mathcal{A}_i)}},
	description={An estimate of $P(\mathcal{A})$ on analysis band $\mathcal{A}_i$}
}

\newglossaryentry{lambdaAk0}{
	type=symbols,
	name={\ensuremath{\lambda_k(\mathcal{A}_0)}},
	description={Eigenvalue associated with $\mathbf{w}_k(\mathcal{A}_0)$}
}

\newglossaryentry{lambdak0}{
	type=symbols,
	name={\ensuremath{\lambda_k^0}},
	description={Equivalent to $\lambda_k(\mathcal{A}_0)$}
}

\newglossaryentry{lambdaki}{
	type=symbols,
	name={\ensuremath{\lambda_k^i}},
	description={Equivalent to $\lambda_k(\mathcal{A}_i)$}
}

\newglossaryentry{numtaper}{
	type=symbols,
	name={\ensuremath{K}},
	description={Number of tapers}
}

\newglossaryentry{numAi}{
	type=symbols,
	name={\ensuremath{I}},
	description={Number of analysis bands $\mathcal{A}_i$}
}

\newglossaryentry{numsubB}{
	type=symbols,
	name={\ensuremath{Q}},
	description={Number of sub-signal bands $\mathcal{B}^q$ in signal band $\mathcal{B}$, for $q=0,1,\ldots,Q-1$}
}

\newglossaryentry{nthx}{
	type=symbols,
	name={\ensuremath{x(t_n)}},
	description={Sample at time $t_n$ of signal $\mathbf{x}$}
}

\newglossaryentry{PA}{
	type=symbols,
	name={\ensuremath{P(\mathcal{A})}},
	description={Integrated spectrum (power) in analysis band $\mathcal{A}$}
}

\newglossaryentry{preci}{
	type=symbols,
	name={\ensuremath{\mathbb{\epsilon}}},
	description={Computation precision of fast NuFFT algorithm}
}

\newglossaryentry{psdnu}{
	type=symbols,
	name={\ensuremath{S(f)}},
	description={Power spectral density of uniformly sampled signal}
}

\newglossaryentry{sig}{
	type=symbols,
	name={\ensuremath{\mathbf{x}}},
	description={Vector of a weakly stationary, band-limited Gaussian process}
}

\newglossaryentry{siglen}{
	type=symbols,
	name={\ensuremath{N}},
	description={Sample size of signal $\mathbf{x}$}
}

\newglossaryentry{subB}{
	type=symbols,
	name={\ensuremath{\mathcal{B}^q}},
	description={$q$-th sub-band of signal band $\mathcal{B}$, $q = 0,1,\ldots,Q-1$. Simplified as $\mathcal{B}$, if context is clear.}
}

\newglossaryentry{tn}{
	type=symbols,
	name={\ensuremath{t_n}},
	description={Sampling time of $n$-th sample, for $n=1,2,\ldots,N$}
}

\newglossaryentry{wk0}{
	type=symbols,
	name={\ensuremath{\mathbf{w}_k^0}},
	description={Simplified notation of $\mathbf{w}_k(\mathcal{A}_0)$, if context is clear.}
}

\newglossaryentry{wk0est}{
	type=symbols,
	name={\ensuremath{\widehat{\mathbf{w}}_k^0}},
	description={Estimate of $\mathbf{w}_k^0$}
}

\newglossaryentry{wki}{
	type=symbols,
	name={\ensuremath{\mathbf{w}_k^i}},
	description={Simplified notation of $\mathbf{w}_k(\mathcal{A}_i)$, if context is clear.}
}

\newglossaryentry{xn}{
	type=symbols,
	name={\ensuremath{x(n)}},
	description={Sample at time $t_n$ of signal $\mathbf{x}$, equivalent to $x(t_n)$. Often refers to a uniformly sampled signal.}
}
\newtheorem{theorem}{Theorem}
\makeglossaries%
\glsdisablehyper

\begin{document}
\title{\rule{480pt}{4pt}
	\algoname---A Computationally Efficient Suboptimal Power Spectrum
	Estimator for Fast Exploration of Nonuniformly Sampled Time Series
	\rule{480pt}{2pt}}

\author{Jie~\textsc{Cui}\textsuperscript{1,2}\,\corrAuthor,
	Benjamin~H.~\textsc{Brinkmann}\textsuperscript{1,2},
	Gregory~A.~\textsc{Worrell}\textsuperscript{1,2,3}}
%\date{}							% Activate to display a given date or no date

\allowdisplaybreaks{}
\maketitle

\hangindent=25pt
\hangafter=1
\indent
\textsuperscript{1}\,Department of Neurology,
\textsuperscript{2}\,Department of Physiology and Biomedical Engineering,
\textsuperscript{3}\,Department of Neural Surgery, Mayo Clinic, 1216 2nd
St.\ SW, Rochester, MN 55902, USA
% need a new paragraph, so keep the next line of space

\hangindent=25pt
\hangafter=1
\indent
\corrAuthor\,Corresponding author: Cui.Jie@mayo.edu.

\begin{center}
	(Published in \textit{Digital Signal Processing}, 18:105834, 2025.
	\textsc{doi}: \href{https://doi.org/10.1016/j.dsp.2025.105834}{10.1016/j.dsp.2025.105834})
\end{center}

\begin{abstract}
	Nonuniformly sampled signals are prevalent in real-world applications.
However, estimating their power spectra from finite samples poses a
significant challenge. The optimal solution---Bronez \gls{gpss} by solving
the associated \gls{gep}---is computationally intensive and thus impractical
for large datasets. This paper describes a fast, nonparametric method:
\gls{m2nufft}, which substantially reduces computational burden while
maintaining statistical efficiency. The algorithm partitions the signal
frequency band into multiple sub-bands.  Within each sub-band, optimal
tapers are computed at a nominal analysis band and shifted to other analysis
bands using the \gls{nufft}, avoiding repeated \gls{gep} computations.
Spectral power within the analysis band is then estimated as the average
power across the taper outputs.  For the special case where the nominal band
is centered at zero frequency, tapers can be approximated via cubic spline
interpolation of \gls{dpss}, eliminating \gls{gep} computation entirely.
This reduces the complexity from $O(\gls{siglen}^4)$ to as low as
$O(\gls{siglen} \log \gls{siglen} + \gls{siglen} \log(1/\gls{preci}))$.
Statistical properties of the estimator, assessed using Bronez \gls{gpss}
theory, reveal that the bias and variance bound of the \gls{m2nufft}
estimator are identical to those of the optimal estimator. Additionally, the
degradation of bias bound indicates deviation from optimality.  Finally, we
propose an extension of Thomson \textit{F}-test to test periodicity in
nonuniform samples.  The estimator's performance is validated through
simulation and real-world data, demonstrating its practical applicability.
The \textsc{Matlab} code of the fast algorithm is available on
GitHub~\cite{CUI2024a}.

\end{abstract}
\newpage

% **************
% main text
% **************
\glsresetall%
% Print the list of abbreviations
\printglossary[type=\acronymtype, title=Table 1. Abbreviations Used Frequently, style=abbtab]

% Print the list of math symbols 
\printglossary[type=symbols, title=Table 2. Mathematical Symbols Used Frequently, style=symbtab]

\section{Introduction}\label{sec:introduction}
Power spectrum estimation is a fundamental tool in a wide variety of
scientific and engineering
disciplines~\cite{ROB1982,KAY1988,PER1993,BRE2014}, including signal
processing, communication, machine learning, physical science, and
biomedical data analysis~\cite{MIT2008,VAS2009}. It allows for the
characterization of the second moments of a time series, elucidating
periodicities, oscillatory behavior, and correlation structures in a signal
process. These attributes are crucial to numerous applications.

Despite its extensive lineage~\cite{ROB1982}, power spectrum estimation
continues to be an active research domain. The primary challenge resides in
estimating the spectrum in a way to minimize bias and ensure statistical
robustness, often from a finite sample of the signal. In many instances,
only a single realization (trial) of the underlying process is available,
making the estimation problem inherently ill-posed~\cite{THO1982,THO2000}.\
The continuous \gls{psd} cannot be directly observed and must be estimated
from discrete, limited samples. This constraint introduces bias and variance
into the estimation, primarily due to spectral leakage caused by time-domain
windowing---equivalent to convolution with the window's Fourier transform in
the frequency domain. Traditional power spectrum estimators such as the
periodogram~\cite{PER1994} are computationally simple but suffer from high
variance in performance. In addition, in many real-world applications, the
signal is often nonuniformly\footnote{Some other terminology has been
interchangeably adopted in literature, such as unevenly, irregularly, and
unequally sampled signal.} sampled. This includes scenarios such as network
packet data transfer~\cite{ENG2007}, laser Doppler
anemometry~\cite{TRO1995,YAR2010}, geophysics~\cite{RAU1997}, atomic clock
analysis~\cite{BER2024},
astronomy~\cite{LOM1978,SCA1982,SPR2020,DOD2023,PAT2024a,PAT2024b}, computer
tomography~\cite{STA1993}, genetics~\cite{LIE2007}, biosensing
optimization~\cite{HOR2024}, and biological
signals~\cite{SAU1994,LAG1995,MIV2023,CUI2024}. Nonuniform sampling often
leads to increased sidelobe leakage and inflated bias in spectral
estimates~\cite{BAB2010}.

This paper focuses on a nonparametric solution to power spectrum estimation
problem, in contrast to parametric methods that assume a specific model of
the time series~\cite{PER1993}. Nonparametric methods are particularly
suitable for rapid, exploratory analysis of large datasets, especially when
the underlying model is unknown. In such cases, Thomson's multitaper method
proves to be a powerful tool~\cite{THO1982}. This method employs the
\gls{dpss}, also known as Slepian sequence~\cite{SLE1978}, denoted as
\gls{dpsskn}, where $k$ indexes the taper (with $1 \le k \le
\gls{numtaper}$), \gls{siglen} is the signal length, and \gls{halfbw} is the
bandwidth. The method computes the \gls{dft} of the uniformly sampled
signal, \gls{xn}, $1 \le n \le \gls{siglen}$, weighted by the \gls{dpss}
taper,
\begin{align*}
    \gls{eigencoeff} = \sum_{n=1}^{\gls{siglen}} \left[x(n) \gls{dpsskn}\right] 
    \mathtt{e}^{-j2 \pi f n / f_s},%
    \numberthis%
    \label{eq:uniform_eigencoefficients}
\end{align*}
where \gls{fs} is the sampling frequency. The power spectrum estimate is
then computed by averaging the squared magnitudes of these eigencoefficients
\gls{eigencoeff}, \gls{eigenspectrum}, known as the $k$-th eigenspectrum,
\begin{align*}
    \hat{P}(f) = \frac{1}{\gls{numtaper}} \sum_{k=1}^{\gls{numtaper}} \gls{eigenspectrum}.%
    \numberthis%
    \label{eq:unifrom_power_spectrum_estimator}
\end{align*}
Adaptive weighting schemes for \gls{eigenspectrum}'s is available to further
improve estimation quality~\cite{THO1982,PER1993}. The multitaper approach
achieves a principled tradeoff between resolution, bias, and variance, and
has been extensively validated across diverse
applications~\cite{TAU1990,BRO1992,PER1993,RIE1994,BAB2014}.

Extending the multitaper scheme to nonuniformly sampled signal is desirable.
However, direct application of the classical multitaper approach is
nontrivial, as estimator performance under nonuniform sampling depends on
more than just frequency resolution. Lepage (2009)~\cite{LEP2009} proposed a
direct generalization of Thomson's original approach~\cite{THO1982},
replacing the DFT with the ``irregular DFT (irDFT)'' and subsequently
replacing the Dirichlet-type kernel with a sampling scheme-dependent,
Hermitian, Toeplitz kernel. This method demonstrated superior performance
compared to competitive multitaper estimates computed from the uniformly
sampled data using interpolation. Springford (2020)~\cite{SPR2020} adapted
the Thomson multitaper method to enhance the estimation from the
Lomb-Scargle (LS) periodogram~\cite{LOM1978,SCA1982}, a technique widely
employed in astronomy. Dodson-Robinson and Haley (2024)~\cite{DOD2024}
evaluated the performance of Chave's \gls{mdss}~\cite{CHA2019} and further
suggested the application of an \textit{F}-test to assess periodicity in
nonuniformly sampled data. Patil et al.~\cite{PAT2024a,PAT2024b} provided
compelling empirical evidence that combining interpolated \gls{dpss} with
the \gls{nufft} can significantly enhance spectral and harmonic analysis of
astrophysical signals.  Additionally, recent developments in compressive
sensing offer alternative strategies for spectrum recovery from randomly
sampled data~\cite{DUA2013}, particularly when the signal exhibits frequency
sparsity. While these approaches have merit in various aspects, a
comprehensive evaluation of their statistical properties in terms of bias,
variance, and optimality has not been adequately evaluated. Moreover, their
computational efficiency has not been systematically addressed.

In contrast to heuristic approaches, the seminal work by Bronez (1985,
1988)~\cite{BRO1985,BRO1988} proposed an optimal estimator based on the
study of the first and second moments of quadratic spectral estimator (see
Section~\ref{sec:bronez_gpss}) for arbitrary sampling times. This method
calculates the optimal weight sequence for each analysis band, known as the
\gls{gpss}, by solving a \gls{gep}. This work established an optimality
criterion for the performance of power spectrum estimators in the general
case of sampling schemes. However, since the number of analysis bands is in
general proportional to \gls{siglen}, and the \gls{gpss} has to be estimated
for each analysis band, the computational cost is prohibitively high for
large \gls{siglen}.

In this study, we have developed a fast algorithm, termed \gls{m2nufft}, by
integrating Thomson's and Bronez's multitaper estimators. Specifically, we
partition the entire signal band, \gls{B}, into multiple sub-bands,
\gls{subB}, $q = 0,1,\ldots,Q-1$, enabling parallel computing. Within each
sub-band, the core idea is to estimate the optimal weight sequence \gls{wk0}
(i.e., the \gls{gpss}) on a nominal analysis band \gls{A0} (see
Section~\ref{sec:m2nufft}).  Rather than solving the computationally
expensive \gls{gep} for other analysis bands \gls{Ai}, for $1 \le i \le
\gls{numAi}-1$, we efficiently shift \gls{wk0} to the \gls{Ai}, using the
\gls{nufft}.  This ensures that for each sub-signal band, we only need to
solve \gls{gep} once. Consequently, the eigencoefficients \gls{eigencoeffAi}
and the integrated spectrum \gls{intspectrum} are readily computed from
\gls{numtaper} eigenspectra \gls{eigenspectrumAi} for each \gls{Ai}. For the
special case where the nominal band \gls{A0} is centered at $f_{c_0} = f_0 =
0$ Hz, further computational reduction may be achieved by interpolating the
uniformly spaced \gls{dpsskn} to the nonuniform grid $t_n$ using cubic
splines~\cite{SPR2020} to approximate the GPSS at $\mathcal{A}_0$,
\gls{wk0est}. In this case, the overall computational complexity is
comparable to that of fast \gls{nufft}, approximately $O(N \log N + N
\log(1/\gls{preci}))$, where \gls{preci} is the precision of
computations~\cite{DUT1993,DUT1996}. 

We evaluated the statistical properties of the proposed method, focusing on
bias, variance, and suboptimality, using the theory developed in Bronez
\gls{gpss}~\cite{BRO1988}. Our findings indicate that the bias and variance
bounds are consistent with those of the optimal method. Additionally, we
suggest that the suboptimality of the fast algorithm may be quantified by
the difference between the approximate and optimal eigenvalues. Furthermore,
an \textit{F}-test has been implemented to assess periodicity in
nonuniformly sampled time series~\cite{THO1982,DOD2023,PAT2024b}.

Importantly, we emphasize that although the main theoretical framework used
in this study was first proposed in the 1980s and 1990s, it remains highly
relevant today. Specifically, the multitaper method of spectral estimation
for arbitrary sampling schemes~\cite{BRO1985,BRO1988,BRO1992,HAN1997}
continues to be a powerful tool for addressing this challenging problem.
This is evidenced by its application in several recent
studies~\cite{CHA2019,BER2024,DOD2023,PAT2024a,PAT2024b,HOR2024,MIV2023},
underscoring its ongoing importance and the necessity for a computationally
efficient implementation.

The reminder of the paper is organized as follows. We provide an overview of
the Bronez GPSS method in Section~\ref{sec:bronez_gpss}, serving as the
theoretical basis for the following developments. In
Section~\ref{sec:m2nufft} we develop the fast M\textsuperscript{2}NuFFT
algorithm\footnote{Dr.~G.M. Eadie of University of Toronto kindly noted
(personal communication) that their group had already adopted a similar name
(i.e.~mtNUFFT) for their methods~\cite{PAT2024a,PAT2024b}. We formulated the
acronym and theory independently.} and evaluate its statistic properties in
the context of Bronez GPSS theory. Section~\ref{sec:taper_subopt} is
dedicated to analyzing taper approximation errors through numerical
experiments.  The performance evaluation of the estimator, which includes
both simulation results and a real-world application, is presented in
Section~\ref{sec:per_eval}. Section~\ref{sec:discussion} offers a broader
discussion, followed by concluding remarks in Section~\ref{sec:summary}. The
\textsc{Matlab} (MathWorks, Natick, MA) code of the fast algorithm
(\gls{m2nufft}, Table~\ref{tb:m2nufft_method}) is publicly available on
GitHub~\cite{CUI2024a}.

\section{Overview of Bronez GPSS Optimal Approach}\label{sec:bronez_gpss}
The \gls{bg} is an extension of the quadratic spectral estimator, developed
to analyze nonuniformly sampled processes~\cite{BRO1985,BRO1988}. It is an
optimal nonparametric method in the sense that it is unbiased in the context
of white noise, and it minimizes the variance and bias bounds for a given
frequency resolution.

Consider \gls{nthx} a weakly stationary, band-limited Gaussian process,
available on a set of arbitrary sampling points \gls{tn}, $1 \le n \le
\gls{siglen}$, where \gls{siglen} is the total number of samples. Instead of
directly estimating the \gls{psd}, \gls{psdnu}, BG estimates the integrated
spectrum (i.e.\ the power), $\gls{PA} = \int_{\gls{A}} \gls{psdnu} \, df$,
contained in an analysis band of interest $\gls{A} = \{f: |f - \gls{fc}| \le
\gls{halfbw}\}$, where \gls{fc} is the center frequency and \gls{halfbw} the
bandwidth. Note that $2\gls{halfbw}$ is the desired frequency resolution. A
complete spectral analysis involves estimating $\gls{PA}$ for a set of
analysis bands, \gls{A}, to cover the entire signal band $\gls{B} = \{f: |f|
\le \gls{fmax}\}$, where \gls{fmax} is presumably the maximum frequency of
the signal~\cite{BRO1988}. The estimator can be expressed as
\begin{align*}
  \hat{P}(\gls{A}) = \frac{1}{K} \mathbf{x}^* \mathbf{Q}(\mathcal{A}) \mathbf{x},%
  \numberthis%
\end{align*}
where $\gls{sig} = [x(t_1), x(t_2), \ldots, x(t_n)]'$, the prime, $'$,
denotes vector transposition, and the asterisk, $*$, denotes complex
conjugate transposition.  The $N \times N$ positive semidefinite Hermitian
weight matrix $\mathbf{Q(\mathcal{A})}$ depends on the analysis band
$\mathcal{A}$. Here, $K \le N$ is the rank of $\hat{P}(\mathcal{A})$. The
weight matrix $\mathbf{Q(\mathcal{A})}$ can be factorized as
$\mathbf{Q(\mathcal{A})} = \mathbf{\Psi}(\mathcal{A})
\mathbf{\Psi}^*(\mathcal{A})$, where $\mathbf{\Psi}(\mathcal{A})$ is an $N
\times K$ matrix. The power spectrum estimator is then given by
\begin{align*}
  \hat{P}(\mathcal{A}) = \frac{1}{K} \sum_{k=1}^{K} \left|\mathbf{w}_k^*(\mathcal{A})
  \mathbf{x}\right|^2,%
  \numberthis%
  \label{eq:gpss_estimator}
\end{align*}
where $\mathbf{w}_k(\mathcal{A})$, $1 \le k \le K$, is the columns of
$\mathbf{\Psi}(\mathcal{A})$.

Assuming that the number of weight sequences, also known as tapers, $K$, is
predetermined, the optimal tapers $\mathbf{w}_k(\mathcal{A})$ are derived
based on the constraints imposed on estimator bias, variance bound, and bias
bound.

\subsection{Bias Constraint}
The estimator, as defined in (\ref{eq:gpss_estimator}), is constrained to be
unbiased when the true spectral density is flat, e.g., $S(f) = 1$. Given the
expectation of the estimate
\begin{align*}
  E\{\hat{P}(\mathcal{A})\} = \int_{\mathcal{B}} \frac{S(f)}{K}
  \sum_{k = 1}^{K} \left|W_k(f)\right|^2 \, df,%
  \numberthis%
\end{align*}
where $W_k(f)$ is the DFT of
$\mathbf{w}_k(\mathcal{A}) = [w_k(t_1), w_k(t_2), \ldots, w_k(t_n)]$,
\begin{align*}
  W_k(f) \triangleq \sum_{n = 1}^{N} w_k(t_n) e^{-j2 \pi f t_n},%
  \numberthis%
  \label{eq:taper_dft}
\end{align*}
to minimize the bias, $E\{\hat{P}(\mathcal{A})\} - P(\mathcal{A})$, the
weight sequences, $\mathbf{w}_k(\mathcal{A})$, must satisfy
\begin{align*}
  \int_{\mathcal{B}} \frac{1}{K} \sum_{k = 1}^{K} \left|W_k(f)\right|^2 \, df
  = \int_{\mathcal{A}} \, df,%
  \numberthis%
\end{align*}
which is equivalent to
\begin{align*}
  \frac{1}{K} \sum_{k = 1}^{K} \mathbf{w}_k^*(\mathcal{A}) \mathbf{R}(\mathcal{B})
  \mathbf{w}_k(\mathcal{A}) = 2f_w,%
  \numberthis%
  \label{eq:bg_bias_contraint}
\end{align*}
where \gls{gmatB} is the GPSS matrix on signal band $\mathcal{B}$. It is an
$N \times N$ positive definite Hermitian matrix, whose elements are
\begin{align*}
  \gls{gmatBnm}
   & = \int_{\mathcal{B}} e^{j2 \pi f(t_n - t_m)} df                      \\
   & = \frac{\sin\left[2 \pi f_{\max}(t_n - t_m)\right]}{\pi(t_n - t_m)}. %
  \numberthis%
  \label{eq:gpss_matrix_b}
\end{align*}

\subsection{Variance Bound}
For a Gaussian process, it can be shown (Appendix B and equation (16)
  in~\cite{BRO1988}) that the variance of the estimator can be bounded above
  by
\begin{align*}
  VAR\{\hat{P}(\mathcal{A})\} \le S_{\max}^2 \cdot
  \mathbf{V}\{\mathbf{w}_1(\mathcal{A}), \ldots, \mathbf{w}_K(\mathcal{A})\},
\end{align*}
where $S_{\max} = \sup_{f \in \mathcal[B]} S(f)$ is the maximum value of the
spectral density, and $\mathbf{V}{\{\mathbf{w}_k(\mathcal{A})\}}_{k = 1}^K$
is the bound factor
\begin{align*}
   & \mathbf{V}\{\mathbf{w}_1(\mathcal{A}), \ldots, \mathbf{w}_K(\mathcal{A})\} \\
   & \quad = \frac{1}{K^2} \sum_{k = 1}^{K} \sum_{l = 1}^{K}                    %
  \left|\int_{\mathcal{B}} {W_k(f)}^* {W_l(f)} \,df\right|^2                    \\
   & \quad = \frac{1}{K^2} \sum_{k = 1}^{K} \sum_{l = 1}^{K}                    %
  \left|\mathbf{w}_k^*(\mathcal{A}) \mathbf{R}(\mathcal{B})
  \mathbf{w}_l(\mathcal{A})\right|^2. %
  \numberthis%
  \label{eq:bg_variance_bound_factor}
\end{align*}
Choosing the weight sequence to minimize
(\ref{eq:bg_variance_bound_factor}), while satisfying
(\ref{eq:bg_bias_contraint}), leads to the sequence normalization requirement
\begin{align*}
  \mathbf{w}_k^*(\mathcal{A}) \mathbf{R}(\mathcal{B}) \mathbf{w}_k(\mathcal{A})
  = 2f_w \quad 1 \le k \le K.%
  \numberthis%
  \label{eq:bg_norm_requirement}
\end{align*}

\subsection{Bias Bound}
By considering only the broad band bias~\cite{THO1982}, the errors due to
frequencies outside analysis band $\mathcal{A}$, an approximate bias of
estimation can be bounded above by
\begin{align*}
  \widehat{BIAS}\{\hat{P}(\mathcal{A})\} \le S_{\max} \cdot
  \mathbf{B}\{\mathbf{w(\mathcal{A})}_1, \ldots, \mathbf{w(\mathcal{A})}_k\},%
  \numberthis%
\end{align*}
where $\mathbf{B}\{\mathbf{w(\mathcal{A})}\}_{k = 1}^K$ is the bias bound
factor (defined in equation (20b) in~\cite{BRO1988}).  Given the
normalization requirement (\ref{eq:bg_norm_requirement}), choosing
$\mathbf{w}_k(\mathcal{A})$ to minimize the bound factor
\begin{align*}
   & \mathbf{B}\{\mathbf{w(\mathcal{A})}_1, \ldots, \mathbf{w(\mathcal{A})}_k\}                                                 \\
   & \quad = \int_{\mathcal{B} - \mathcal{A}} \frac{1}{K} \sum_{k = 1}^{K} \left|W_k(f)\right|^2 \,df \\
   & \quad = \frac{1}{K} \sum_{k = 1}^{K} \mathbf{w}_k^*(\mathcal{A})
  [\mathbf{R}(\mathcal{B}) - \mathbf{R}(\mathcal(A))] \mathbf{w}_k(\mathcal{A}),%
  \numberthis%
\end{align*}
where the cut integral is defined as $\int_{\mathcal{B} - \mathcal{A}} =
  \int_{\mathcal{B}} - \int_{\mathcal{A}}$. This results in the GEP
\begin{align*}
  \mathbf{R}(\mathcal{A}) \mathbf{w}_k(\mathcal{A}) = \lambda_k^{\mathcal{A}}
  \mathbf{R}(\mathcal{B})
  \mathbf{w}_k(\mathcal{A}), \quad 1 \le k \le K,%
  \numberthis%
  \label{eq:gep}
\end{align*}
where $\mathbf{R}(\mathcal{B})$ is the GPSS matrix on the signal band shown
in (\ref{eq:gpss_matrix_b}) and $\mathbf{R}(\mathcal{A})$ is the GPSS matrix
on the analysis band
\begin{align*}
  \gls{gmatAnm}
   & = \int_{\mathcal{A}} e^{j2 \pi f(t_n - t_m)} df                                           \\
   & = \frac{\sin\left[2 \pi f_w(t_n - t_m)\right]}{\pi(t_n - t_m)}e^{j 2\pi f_c (t_n - t_m)}. %
  \numberthis%
  \label{eq:gpss_matrix_a}
\end{align*}

The GEP (\ref{eq:gep}) has $N$ independent solutions
$\{\lambda_k^{\mathcal{A}}, \mathbf{w}_k(\mathcal{A})\}$, $1 \le k \le N$,
for the analysis band $\mathcal{A}$. The weight sequences corresponding to
the $K$ largest eigenvalues are chosen to minimize the bound factors.

The computation of the GEP (\ref{eq:gep}) requires $O(N^3)$ operations for
each analysis band of interest. In general, the number of bands is
proportional to the number of samples, $N$, and thus makes the total
computational load on the order of $O(N^4)$. The computational demand may be
impractical when $N$ is large.

\subsection{Analysis in a Sub-Signal Band}\label{sec:sub_band}

In the preceding discussion, the signal band \gls{B} was assumed to span the
entire frequency range of the signal, which is typically the case in
practice. However, as discussed in Section~\ref{sec:m2nufft}, it can be
advantageous to consider sub-signal bands for localized analysis and
parallel computation.  A sub-band is defined as $\gls{subB} = \{f:
f_{\min}^q \leq |f| \leq f_{\max}^q \}$, where $f_{\min}^q \geq 0$ and
$f_{\max}^q \leq f_{\max}$ denote the lower and upper bounds of \gls{subB},
and $q = 0, 1, \ldots, \gls{numsubB}-1$ indexes the \gls{numsubB} disjoint
sub-bands, collectively covering the entire signal band, $\gls{B} =
\bigcup_{q=0}^{\gls{numsubB}-1} \gls{subB}$.  In this paper, we assume that
the partition of the signal band into \gls{numsubB} sub-band \gls{subB} is
pre-determined (Table~\ref{tb:m2nufft_method}, but see the discussion in
Section~\ref{sec:discussion}).

From the \gls{gpss} formulation in equation~(\ref{eq:gpss_matrix_b}) and the
Bronez's theorem stated in~\cite{BRO1988}, the \gls{gpss} matrix
$\mathbf{R}(\mathcal{B}^q)$ corresponding to each sub-band \gls{subB}
remains a positive definite Hermitian matrix (\ref{app:pos_def_herm}), whose
elements are 
\begin{align*}
  \mathbf{R}(\gls{subB}| n, m)
   & = \int_{-f_{\max}^q}^{f_{\max}^q} e^{j2 \pi f(t_n - t_m)} df        %
  - \int_{-f_{\min}^q}^{f_{\min}^q} e^{j2 \pi f(t_n - t_m)} df           \\
   & = 2\frac{\cos\left[\pi (f_{\max}^q + f_{\min}^q) (t_n - t_m)\right] %
    \sin\left[\pi (f_{\max}^q - f_{\min}^q) (t_n - t_m)\right]} %
  {\pi(t_n - t_m)}, %
  \numberthis%
  \label{eq:gpss_matrix_b_q}
\end{align*}
where $f_{\min}^q < f_{\max}^q$.  The \gls{gep} in equation~(\ref{eq:gep})
can be solved independently for each \gls{subB} to obtain the optimal weight
sequences $\mathbf{w}_k(\mathcal{A}^q)$.  Importantly, the constraints
governing bias control, normalization, and optimality---namely,
equations~(\ref{eq:bg_bias_contraint}), (\ref{eq:bg_norm_requirement}) and
(\ref{eq:gep})~extend naturally to each sub-band \gls{subB}.  The full
frequency band \gls{B} can thus be viewed as a special case, where
$f_{\max}^q = f_{\max}$, $f_{\min}^q = 0$, and $\gls{numsubB} = 1$.

\section{Multiband-Multitaper Nonuniform Fast Fourier Transform}\label{sec:m2nufft}
In this section, we present the core structure of the Multiband-Multitaper
Nonuniform Fast Fourier Transform (\algoname) method, designed for rapid
power spectrum estimation in nonuniformly sampled time series. The
derivation of this method assumes that the series follows a
weakly-stationary, band-limited Gaussian process, similar to previously
introduced methods~\cite{BRO1985,BRO1988}. The number of weight sequences,
or tapers, denoted as \gls{numtaper}, is predetermined and correlates with
the properties of the tapers obtained. We will evaluate the statistical
performance of the estimator based on bias measure, variance bound, and
sidelobe leakage. The quantification of leakage may serve as a measure of
suboptimality.

\subsection{Multiband Partition of Signal Band}
To reduce computational load and enable parallelization, we partition the
full signal band \gls{B} into \gls{numsubB} non-overlapping sub-bands,
\gls{subB}, $q = 0,1, \ldots \gls{numsubB}-1$.  Each sub-band contains a
group of analysis bands (c.f., Section~\ref{sec:sub_band}). This partition
is particularly effective when $\gls{numsubB} \ll \gls{siglen}$, as it
requires the \gls{gep} be solved only once per sub-band, significantly
lowering computational cost. Moreover, since computation across sub-bands
are independent, the algorithm is naturally suited for parallel computing
architecture.

Within each sub-band \gls{subB}, we define a set of analysis bands,
$\gls{A}_i^q \subset \gls{B}^q$, each characterized by a center frequency
$\gls{fci}^q$ and bandwidth $\gls{halfbw}^q$,
\begin{align*}
  \mathcal{A}_i^q = \left\{f: |f - f_{c_i}^q| \le f_w^q\right\},
  \quad 0 \le i \le I^q-1,\numberthis
\end{align*}
where $I^q$ is the number of analysis bands in sub-band ${\gls{subB}}$.
These bands are identical, differing only by a frequency shift.  The
frequency resolution within each analysis band is $2\gls{halfbw}^q$, and the
boundaries satisfy $f_{\min}^q = f_{c_{\min}} - f_w^q$, $f_{\max}^q =
f_{c_{\max}} + f_w^q$. The full-signal band case is recovered when
$\gls{numsubB} = 1$, i.e., when the entire frequency range is treated as a
single sub-band.

\begin{figure}[!t]
    \centering
    \includegraphics[width=4.5in]{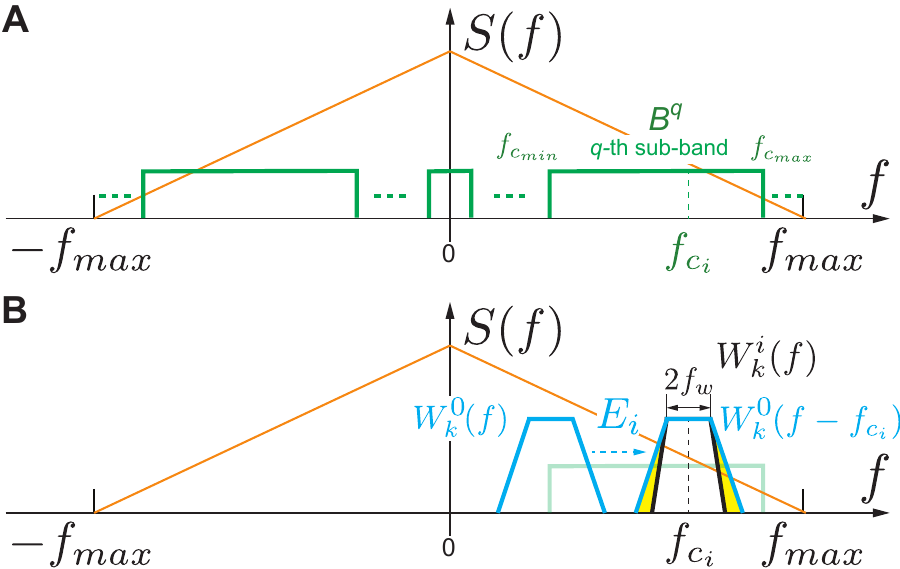}
    \caption{\textbf{Frequency-Domain Shift of Optimal Weighting Sequence.}
        Panel (\textbf{A}) illustrates the partitioning the signal band,
        defined as $\mathcal{B} = \{f: |f| \le f_{\max}\}$, into multiple
        non-overlapping segments or sub-band. Each sub-band comprises a
        group of analysis bands, such as the \textit{q}-th sub-band
        containing bands centered at frequencies $f_{c_i}$, ranging from
        $f_{c_{min}}$ to $f_{c_{max}}$. Panel (\textbf{B}) depicts the
        frequency-domain shift of the optimal weighting sequence, $W_k^0(f)$
        (shown as a light blue trapezium), within the \textit{q}-th band
        group. The shift moves the sequence from its original center at $f =
        f_{c_0}^q$ to a target analysis band centered at $f_{c_i}^q$. The
        bandwidth of each analysis band is $2f_w^q$ (superscript $q$ omitted
        for clarity thereafter). The operator $E_i$ denotes the frequency
        shift.  The dark trapezium represents the frequency-domain transform
        of the optimal weighting sequence, $W_k^i(f)$, centered at $f_{c_i}$
        as per Bronez GPSS approach~\cite{BRO1985,BRO1988}.  Yellow shading
        highlights the sidelobe leakage difference---indicating
        suboptimality---between the optimal $W_k^i(f)$ and the shifted
        version $W_k^0(f - f_{c_i})$. The orange triangle represents the
        signal power spectrum, $S(f)$.}%
    \label{fig:shift_eigenvec}
\end{figure}

As discussed in Section~\ref{sec:bronez_gpss}, the primary computational
bottleneck in \gls{bg} method lies in the adaptive estimation of the tapers
for each analysis band. To alleviate this computational burden, we propose
computing the optimal tapers ${\{{}^q\mathbf{w}_k^0\}}_{k = 1}^K$ only once
at a nominal band $\mathcal{A}_0^q$ centered at $f_{c_0}^q$, and then
shifting these tapers to all other analysis bands $\gls{A}_i^q$ within the
sub-band \gls{subB} using the \gls{nufft} (see
Figure~\ref{fig:shift_eigenvec}). This approach avoids repeated \gls{gep}
solutions.

Since the analysis procedure is identical for each sub-band $\gls{subB}$, we
omit the subscript $q$ in the following derivations for clarity, unless
otherwise specified.

% one-column width = 5in; two-column width = 3.45in

\subsection{Multitaper Nonuniform FFT (MTNuFFT) Estimator}
In each sub-band $\mathcal{B}$, we designate $\mathcal{A}_0$ centered at
$f_{c_0}$ as the nominal analysis band. The corresponding optimal tapers,
denoted as $\{\mathbf{w}_k^0\}_{k = 1}^K$, are determined by solving the GEP
(\ref{eq:gep}) at $f_{c_0}$,
\begin{align*}
  \mathbf{R}(\mathcal{A}_0)\mathbf{w}_k^0 = \lambda_k^0 \mathbf{R}(\mathcal{B})
  \mathbf{w}_k^0, \quad 1 \le k \le K,%
  \numberthis%
  \label{eq:gep_at_A0}
\end{align*}
where $\{\lambda_k^0, \mathbf{w}_k^0\}$ represents the $k$-th pair of
eigenvalue and eigenvector for $\mathcal{A}_0$. The elements of the $N
  \times N$ positive definite Hermitian matrix $\mathbf{R}(\mathcal{A}_0)$ are
given by (\ref{eq:gpss_matrix_a})
\begin{align*}
  \mathbf{R}(\gls{A0}| n, m) & = \int_{\mathcal{A}_0} \mathtt{e}^{j2 \pi f(t_n - t_m)} df                                                    \\
                                  & = \frac{\sin\left[2 \pi f_w(t_n - t_m)\right]}{\pi(t_n - t_m)} \cdot \mathtt{e}^{j 2\pi f_{c_0} (t_n - t_m)}. %
  \numberthis
\end{align*}
We define the frequency shifting operator \gls{Ei} as
\begin{align*}
  \gls{Ei} = \begin{bmatrix}
               \mathtt{e}^{j2 \pi \Delta f_i t_1} & 0                                  & \cdots & 0                                  \\
               0                                  & \mathtt{e}^{j2 \pi \Delta f_i t_2} & \cdots & 0                                  \\
               \vdots                             & \vdots                             & \ddots & \vdots                             \\
               0                                  & 0                                  & \cdots & \mathtt{e}^{j2 \pi \Delta f_i t_N}
             \end{bmatrix},%
  \numberthis%
  \label{eq:freq_shift_op}
\end{align*}
where $\Delta f_i = f_{c_i} - f_{c_0}$. Note that if we choose the center
frequency of the nominal band $\mathcal{A}_0$ at $f_{c_0} = 0$, then $\Delta
  f_i = f_{c_i}$. To void solving computationally costed GEP problem at
$\mathcal{A}_i$, we may approximate the eigenvector $\mathbf{w}_k^i$ with
$E_i \mathbf{w}_k^0$, which is shifted from $\mathbf{w}_k^0$, so that
$J_k(\mathcal{A}_i) = {\mathbf{w}_k^i}^* \mathbf{x}$, known as the
eigencoefficients~\cite{PER1993}, may be estimated as
\begin{align*}
  \hat{J_k}(\mathcal{A}_i) & = {\left(E_i \mathbf{w}_k^0\right)}^* \mathbf{x}                              \\
                           & = \sum_{n=1}^N \left[{w_k^0}^*(t_n) x(t_n)\right] e^{-j2 \pi \Delta f_i t_n}.
  \numberthis%
  \label{eq:eigencoefficients}
\end{align*}
The power spectrum estimator of the integrated spectrum over analysis band
$\mathcal{A}_i$ is then given by
\begin{align*}
  \hat{P}(\mathcal{A}_i) = \frac{1}{K} \sum_{k=1}^K \left|\hat{J_k}(\mathcal{A}_i)\right|^2.%
  \numberthis%
  \label{eq:power_spectrum_estimator}
\end{align*}
The eigencoefficients (\ref{eq:eigencoefficients}) is typically implemented
using Nonuniform FFT (NuFFT)~\cite{POT2017,DUT1996}.  However, note that in
sub-band $\mathcal{B}^q$, we use the relative frequency points $\Delta
  f_{c_i}$ in calculating NuFFT.  The desired frequency points are $f_{c_i} =
  f_{c_0} + \Delta f_{c_i}$.  As for the special case, when the central
frequency of $\mathcal{A}_0$ is $f_{c_0} = 0$, $\Delta f_{c_i} = f_{c_i}$.
A relevant work to the NuFFT is the Nonuniform Discrete Fourier Transform
(NDFT)~\cite{BAG1996} of a time series, which is defined as samples of its
$z$-transform evaluated at distinct points located nonuniformly on the unit
circle in the $z$-plane.

\subsection{Bias Measure}
We began by evaluating the performance of the estimator
(\ref{eq:power_spectrum_estimator}) in terms of bias. This assessment was
carried out under the condition that the signal is white, meaning the true
spectral density is flat, as previously employed in Bronez GPSS
approach~\cite{BRO1988}. Specifically, we consider the case where $S(f) =
  1$. The expectation of the estimator can be expressed as
\begin{align*}
  \mathbf{E}\left[\hat{P}(\mathcal{A}_i)\right]
   & = \int_{\mathcal{B}} S(f) \frac{1}{K} \sum_{k=1}^K \left| W_k^i(f) \right|^2 df      \\
   & = \int_{\mathcal{B}} \frac{1}{K} \sum_{k=1}^K \left| W_k^0(f) \right|^2 df           \\
   & = \frac{1}{K} \sum_{k=1}^K {\mathbf{w}_k^0}^* \mathbf{R}(\mathcal{B}) \mathbf{w}_k^0 \\
   & = \mathbf{E}\left[\hat{P}(\mathcal{A}_0)\right],
  \numberthis
  \label{eq:bias_measure}
\end{align*}
where we have replaced the DFT $\mathbf{W}_k^i(f)$ with $\mathbf{W}_k^0(f)$
according to the algorithm, and $\mathbf{w}_k^0$ is optimal at
$\mathcal{A}_0$. Clearly, we have
$\mathbf{E}\left[\hat{P}(\mathcal{A}_0)\right] = 2f_w$ by satisfying the
normalization requirement (\ref{eq:bg_norm_requirement}).  The bias of the
estimator is then
\begin{align*}
  \text{BIAS}\{\hat{P}(\mathcal{A}_i)\} = \mathbf{E}\left\{\hat{P}(\mathcal{A}_i)\right\}
  - \int_{f_{c_i}-f_w}^{f_{c_i}+f_w} S(f) \,df = 0.%
  \numberthis
\end{align*}

\subsection{Variance Bound}
From (\ref{eq:bg_variance_bound_factor}), for Gaussian process, the bound
factor may be written as
\begin{align*}
  \mathbf{V}(\widehat{\mathbf{w}}_1^i, \ldots, \widehat{\mathbf{w}}_K^i)
   & = \mathbf{V}(E_i \mathbf{w}_1^0, \ldots, E_i \mathbf{w}_K^0) \\
   & = \frac{1}{K^2} \sum_{k = 1}^K \sum_{l = 1}^K
  \left|\int_{\mathcal{B}} {W_k^0}^*(f-f_{c_i}) \times {W_l^0}(f-f_{c_i}) \,df\right|^2.         %
  \numberthis%
  \label{eq:bound_factor}
\end{align*}
If the frequency center of the tapers are not near to the boundary of signal
band $\mathcal{B}$, for instance, $f_{c_i}$ and $\pm f_{\max}$ are separated
by at least $2f_w$, (\ref{eq:bound_factor}) may be approximated as
\begin{align*}
  \mathbf{V}(\widehat{\mathbf{w}}_1^i, \ldots, \widehat{\mathbf{w}}_K^i)
   & \approx \frac{1}{K^2} \sum_{k = 1}^K \sum_{l = 1}^K
  \left|\int_{\mathcal{B}} {W_k^0}^*(f){W_l^0}(f) \,df\right|^2         \\
   & = \frac{1}{K^2} \sum_{k=1}^K \sum_{l=1}^K \left|{\mathbf{w}_k^0}^*
  \mathbf{R}(\mathcal{B}) \mathbf{w}_l^0\right|^2                       \\
   & = \frac{(2f_w)^2}{K},
  \numberthis%
  \label{eq:bound_factor_approx}
\end{align*}
which is identical to the bound factor of the optimal approach.

\subsection{Sidelobe Leakage and Suboptimality}\label{sec:subopt_measure}
As we shift the optimal eigenvectors (tapers) from the nominal analysis band
$\mathcal{A}_0$ to $\mathcal{A}_i$, rather than using the optimal
eigenvectors at the designated analysis band, it becomes crucial to
understand the deviation from the optimal solution. As previously discussed,
the bias measure and variance bound factor match the optimal ones, provided
that the analysis band is not in proximity to $\pm f_{\max}$. We now
consider the difference in bias bound factor between the optimal and our
proposed solutions. We utilize this difference as a metric to indicate
suboptimality of the fast algorithm (Fig.~\ref{fig:shift_eigenvec}).

Using the identity $\mathbf{R}(\mathcal{A}_0) = E_i^*
  \mathbf{R}(\mathcal{A}_i) E_i$, where $\mathcal{A}_i = \mathcal{A}_0 +
  2\pi (f_{c_i}-f_{c_0})$ (c.f., (3.42) in~\cite{BRO1985}), the bias bound
factor can be expressed as
\begin{align*}
  \mathbf{B}(\widehat{\mathbf{w}}_1^i, \ldots, \widehat{\mathbf{w}}_K^i)
   & = \int_{\mathcal{B}} \frac{1}{K} \sum_{k=1}^K \left|W_k^q(f \
  - f_{c_i})\right|^2 \,df - \frac{1}{K} \sum_{k=1}^K {\left(E_i \mathbf{w}_k^q\right)}^* \mathbf{R}(\mathcal{A}_i) {\left(E_i \mathbf{w}_k^q\right)}                                                                  \\
   & \approx \int_{\mathcal{B}} \frac{1}{K} \sum_{k=1}^K \left|W_k^q(f)\right|^2 \,df                                                                                                                                \
  - \frac{1}{K} \sum_{k=1}^K {\mathbf{w}_k^0}^* \mathbf{R}(\mathcal{A}_0) \mathbf{w}_k^0                                                                                                                               \\
   & = \frac{1}{K} \sum_{k=1}^K {\mathbf{w}_k^0}^* [\mathbf{R}(\mathcal{B}) - \mathbf{R}(\mathcal{A}_0)] \mathbf{w}_k^0                                                                                                \\
   & = \frac{2 f_w}{K} \sum_{k=1}^K (1-\lambda_k^0).                                                                                                                                                                   %
  \numberthis
\end{align*}
We once again assume that $\mathcal{A}_i$ is not near the boundary of
$\mathcal{B}$. The absolute value of bound factor difference can now be
readily seen as
\begin{align*}
  \vert\Delta\mathbf{B}_i\vert & \triangleq
  \vert\mathbf{B}(\widehat{\mathbf{w}}_1^i, \ldots, \widehat{\mathbf{w}}_K^i)
  - \mathbf{B}(\mathbf{w}_1^i, \ldots, \mathbf{w}_K^i)\vert                                              \\
                               & = \frac{2 f_w}{K} \vert \sum_{k=1}^K (\lambda_k^i-\lambda_k^0) \vert    \\
                               & \leq \frac{2 f_w}{K} \sum_{k=1}^K \vert \lambda_k^i-\lambda_k^0 \vert . %
  \numberthis%
  \label{eq:bias_bound_diff}
\end{align*}
The difference (\ref{eq:bias_bound_diff}) suggests that
\begin{align*}
  \gls{E} = \frac{1}{K} \sum_{k=1}^K \vert \lambda_k^i-\lambda_k^0 \vert %
  \numberthis%
  \label{eq:subopt_measure}
\end{align*}
may serve as a measure of deviation from the optimal case. Clearly, $\gls{E}
  \in [0,1]$ due to the eigenvalue condition $0 \le \lambda_k^0, \lambda_k^i
  \le 1$.

\subsection{Thomson \textit{F}-test for Nonuniform Signal}

Statistical tests are often employed to ascertain the periodicity in signal.
When a spectral peak is observed, it's crucial to determine if its magnitude
significantly exceeds what could arise by chance. The Thomson
\textit{F}-test~\cite{THO1982} serves as an effective tool for detecting
spectral lines in colored noise (i.e., mixed spectrum), including biological
signals~\cite{MIT1999,MIT2008}.

The \textit{F}-statistic for the nonuniformly sampled time series may be
formally computed from the eigencoefficients (\ref{eq:eigencoefficients}).
Assuming $2f_{c_i} > f_w$, the \textit{F}-statistic can be derived as
(c.f., pp. 496--500 in~\cite{PER1993})
\begin{align*}
  F_s = \frac{|\hat{C}_i|^2 (K-1) \sum_{k=1}^K\left[W_k^0(0)\right]^2}
  {\sum_{k = 1}^K \left|J_k(\mathcal{A}_i) - \hat{C}_i W_k^0(0)\right|^2},%
  \numberthis%
  \label{eq:f_stat}
\end{align*}
where $W_k^0(0)$ is the NuFFT (\ref{eq:eigencoefficients}) of
$\mathbf{w}_k^0$ at $\Delta f_i = 0$ (i.e., $f_{c_i} = f_{c_0}$), which is
simply the summation of taper weights, $W_k^0(0) = \sum_{n = 1}^N
  w_k^0(t_n)$. $\hat{C}_i$ is the estimated amplitude at $f_{c_i}$, calculated
as
\begin{align*}
  \hat{C}_i = \frac{\sum_{k = 1}^K J_k(\mathcal{A}_i) W_k^0(0)}
  {\sum_{k = 1}^K \left[W_k^0(0)\right]^2}.%
  \numberthis
\end{align*}
The statistic in (\ref{eq:f_stat}) follows an \textit{F}-distribution, $F_s
  \sim F(2, 2K-2)$, with 2 and $2K-2$ degrees of freedom. The critical value
$F_{\alpha}$ for a given level $\alpha = 1 - p$ can be found from the
inverse \textit{F}-distribution. As a general guideline, it is
recommended~\cite{THO1982,PER1993} to set the \textit{p}-value at the
Rayleigh frequency $1/N$, where $N$ denotes the number of sample points.

Other studies proposed some similar statistical tests for spectral lines in
nonuniformly sampled time series~\cite{CHA2019,DOD2024,PAT2024b}. However,
these methods generally do not normalize the tapers according to
(\ref{eq:bg_norm_requirement}), which is essential for ensuring the energy
conservation.

\subsection{Computational Cost and \texorpdfstring{\gls{m2nufft}}~Algorithm}

The computation of the optimal taper \gls{wk0} necessitates the solution of
the \gls{gep} at \gls{A0}, as defined in~(\ref{eq:gep_at_A0}). This step
incurs a computational cost of $O(N^3)$, which is significant but performed
only once per sub-band \gls{subB}, unlike Bronez's original GPSS method,
where the \gls{gep} must be solved for every analysis band. The transition
of \gls{wk0} to other analysis bands relies on the \gls{nufft}, which
demands $O(N \log N + N \log(1/\gls{preci}))$ arithmetic
operations~\cite{DUT1993,DUT1996}, where $\gls{preci}$ is the precision of
computation, without repeated \gls{gep} solutions.

\begin{table}[!t]
   \centering
   \caption{M\textsuperscript{2}NuFFT Algorithm of Spectral Estimation}%
   \label{tb:m2nufft_method}
   \renewcommand{\arraystretch}{0.8}
   \begin{tabular}{p{0.1in}p{4.5in}p{0.1in}}
      \hline
      \hline
      \textbf{1}
       & Define sampling points $t_n, 1 \le n \le N$, time series samples
      $x(t_n), 1 \le n \le N$, signal band $\mathcal{B} = \{f: |f| \le
         f_{\max}\}$, and the number of tapers $K$.
       &                                                                                            \\
      % \hline
      \textbf{2}
       & Define the \underline{multiband} $\mathcal{B}^q = \{f: f_{min}^q \le |f| \le f_{max}^q\}$, %
      nominal band $\mathcal{A}_0^q$ centered at $f_{c_0}^q$, and the half
      bandwidth $f_w^q$, for $q = 0,1, \ldots, \gls{numsubB}-1$. &
      \\
      % \hline
      \textbf{3}
       & \textbf{FOR} $q = 0,1, \ldots, \gls{numsubB}-1$ \textbf{DO:}                               %
       &                                                                                            \\
      % \hline
       & \textbf{IF} $f_{c_0}^0 = f_0 = 0$ \textbf{DO:}                                             %
       &                                                                                            \\
      % \hline
       & Derive ${}^q \widehat{\mathbf{w}}_k^0 = \{{^q }w_k^0(t_1), \ldots,                         %
         {}^q w_k^0(t_N)\}$ by interpolation:
       &                                                                                            \\
      % \hline 
       & (\textbf{a}) Compute DPSS on a uniform sampling grid, denoted as
      $v_n^{(k)}(N,f_w)$, where $k$ is the order of the sequence. The grid
      interval is determined by the average inter-sample-interval
      $\overline{\Delta}_t = (t_N - t_1)/N$.
       &                                                                                            \\
      % \hline
       & (\textbf{b}) The taper weights $w_k^0(t_n)$ at intermediate
      points corresponding to the nonuniform times $t_n$ are obtained by
      interpolation using a cubic spline.
       &                                                                                            \\
      % \hline
       & (\textbf{c}) Normalize the taper weights (eigenvector) such that
       &                                                                                            \\
      % \hline
       & \multicolumn{1}{c}{$\displaystyle{\widehat{\mathbf{w}}_k^{0^*}                             %
               \mathbf{R}(\mathcal{B}) \widehat{\mathbf{w}}_k^0 = 2f_w}, \, 1 \le k \le K$.}
       & \multicolumn{1}{r}{(\ref{eq:bg_norm_requirement})}                                         \\
      % \hline
       & \textbf{ELSE DO:}                                                                          %
       &                                                                                            \\
      % \hline
       & Find ${}^q\widehat{\mathbf{w}}_k^0 = \{{}^q w_k^0(t_1), \ldots,                            %
         {}^q w_k^0(t_N)\}$ by solving the \gls{gep}:\@
       &                                                                                            \\
      % \hline
       & \multicolumn{1}{c}{$\displaystyle{                                                         %
               \mathbf{R}(\mathcal{A}_0^q) \cdot {}^q\mathbf{w}_k^0(\mathcal{A}_0^q)%
               = \lambda_k^{\mathcal{A}_0^q}
               \mathbf{R}(\mathcal{B}^q) \cdot
               {}^q\mathbf{w}_k(\mathcal{A}_0^q), \quad 1 \le k \le K,%
            }$}                                                                            %
       & \multicolumn{1}{r}{(\ref{eq:gep})}                                                         \\
      % \hline
       & and normalize ${}^q\widehat{\mathbf{w}}_k^0$.
       & \multicolumn{1}{r}{(\ref{eq:bg_norm_requirement})}                                         \\
      % \hline
       & \textbf{END IF}                                                                            %
       &                                                                                            \\
      % \hline
      \textbf{4}
       & Calculate the eigencoefficients by \underline{\gls{nufft}:}\
       &                                                                                            \\
      % \hline
       & \multicolumn{1}{c}{$\displaystyle J_k(\mathcal{A}_i^q) = \sum_{n=1}^N                      %
            \left[{}^q w_k^{0*}(t_n) x(t_n)\right] e^{-j2 \pi \Delta f_i t_n}$,}
       &                                                                                            \\
      % \hline
       & \multicolumn{1}{c}{$1 \le k \le K, \, 0 \le i \le I-1$.}                                   %
       & \multicolumn{1}{r}{(\ref{eq:eigencoefficients})}                                           \\
      % \hline
      \textbf{5}
       & Compute the \underline{multitaper} estimator of integrated spectrum:
       &                                                                                            \\
      % \hline
       & \multicolumn{1}{c}{$\displaystyle{\hat{P}(\mathcal{A}_i) =                                 %
               \frac{1}{K} \sum_{k=1}^K \left|J_k(\mathcal{A}_i)\right|^2, \,
               0 \le i \le I-1}$.}
       & \multicolumn{1}{r}{(\ref{eq:power_spectrum_estimator})}                                    \\
      % \hline
       & \textbf{END FOR}                                                                           %
       &                                                                                            \\
      \hline
   \end{tabular}
\end{table}

Further computational savings are possible when the nominal analysis band
\gls{A0} is centered at $f_{c_0} = 0$ Hz. In this special case, the optimal
taper \gls{wk0} may be approximated using the conventional \gls{dpss}
\gls{dpsskn}, which are efficiently computable with fast
algorithms~\cite{PER1993,KAR2019}. Given the sufficient regularity of the
\gls{dpsskn}, it is advantageous to interpolate the uniformly sampled
\gls{dpss} to nonuniform grid using a cubic spline~\cite{SPR2020} to
circumvent the computation of \gls{gep}.  Consequently, the overall
computational cost of the fast \gls{m2nufft} algorithm approximates that of
\gls{nufft}.

Instead of normalizing the power of \gls{wk0est} to unity, we adhere to
normalization requirements in equation~(\ref{eq:bg_norm_requirement}). This
normalization is of theoretical importance to ensure energy conservation
when transforming between time and frequency domains. While alternative
normalization methods---such as $L^2$-norm normalization of the interpolated
\gls{dpss}~\cite{SPR2020,PAT2024a}---may introduce a constant bias in the
spectrum, the bias is often negligible in practice, as most applications
focus on relative changes of power spectrum rather than absolute values.

The complete \gls{m2nufft} algorithm for spectral estimation in nonuniformly
sampled time series is summarized in Table~\ref{tb:m2nufft_method}. Notably,
the method proposed by Patil et al.~\cite{PAT2024a} may be viewed as a
special case of \gls{m2nufft}, where $\gls{numsubB} = 1$, the signal band
\gls{B} covering the entire frequency band, and $f_{c_0} = 0$. In this
scenario, the computational complexity is effectively that of \gls{nufft}.
At the other extreme, when $Q = I$, where $I$ is the total number of
analysis bands, the \gls{m2nufft} algorithm becomes equivalent to the
original Bronez \gls{gpss} method with full frequency-dependent taper
estimation.

Finally, the multiband structure described in Section~\ref{sec:sub_band} is
inherently compatible with parallel computing architectures, enabling
further reductions in computational time.

\section{Taper Suboptimality Analysis}\label{sec:taper_subopt}
As previously discussed within the framework of Bronez GPSS~\cite{BRO1988},
the suboptimality of the \algoname algorithm primarily arises from two
factors:

\begin{enumerate}
    \item \textbf{Band mismatch:} The discrepancy between the optimal tapers
          $\mathbf{w}_k^0$, $k = 1, \ldots, K$, defined at the nominal
          analysis band $\mathcal{A}_0$, and the optimal tapers
          $\mathbf{w}_k^i$ corresponding to the designated analysis band
          $\mathcal{A}_i$, for $i = 1, \ldots, I-1$.     
    \item \textbf{Interpolation error:} When an interpolated DPSS, denoted
          $\hat{\mathbf{w}}_k^0$, is used to approximate the optimal tapers
          at $\mathcal{A}_0$ (where $f_c = 0$) for computational efficiency,
          deviations between $\hat{\mathbf{w}}_k^0$ and the true optimal
          tapers $\mathbf{w}_k^0$ may contribute additional suboptimality.
\end{enumerate}

Due to the complexity introduced by arbitrary sampling schemes, a
closed-form analytical characterization of these discrepancies is beyond the
scope of this work. Instead, we resort to numerical experiments to assess
the suboptimality introduced by the \algoname\ algorithm.  Specifically, we
computed the taper errors between \algoname-generated tapers and the
corresponding optimal tapers under four representative sampling scenarios:
Uniformly Sampling, Jittering Sampling, Missing Data, and Arithmetic
Sampling. The frequency domain is normalized to the interval 0--0.5 Hz,
assuming a maximum signal frequency of $f_{\text{max}} = 0.5$ Hz. All
simulated signals have a fixed duration of $T = 50$ seconds.

We adopted the sampling schemes implemented in~\cite{BRO1985} to generate 50
timestamps.  For uniform sampling, the samples were acquired at one-second
interval, denoted as $t_n = n$, for $1 \le n \le 50$. To construct the
jittering timestamps~\cite{BAR1963,BRE2014}, the sampling time was defined
as $t_n = n + z_n$, where the jittered displacement process $z_n$ was drawn
from a Gaussian white noise (GWN) distribution with zero mean and a standard
deviation of 0.1 seconds. The average sampling rate, or
intensity~\cite{BRE2014}, was set to 1 sample per second. For the
missing-data sampling scheme, data were initially sampled at time points
$t_n = 5n/6$, for $1 \le n \le 60$, followed by the random omission of 10
samples to simulate data loss. The fourth set of timestamps employed
arithmetic sampling, defined as $t_n = 1 + a(n-1) + b(n-1)(n-2)/2$, for $1
\le n \le 50$, where $a$ and $b$ are random variables governed by a control
probability $\rho$~\cite{BRO1985}.

For multitaper spectral estimation in~\algoname, we selected a bandwidth
$f_w = 0.05$ Hz, yielding a frequency resolution of 0.1 Hz and a
time-bandwidth (TW) of 2.5\footnote{TW is conventionally defined as the
product of the signal duration $T$ and the bandwidth $f_w$, rather than the
full bandwidth (frequency resolution) $2f_w$.}. A total of $K = 4$ tapers
were used for the analysis. For the Bronez \gls{gpss} methods (BGFixed and
BGAdaptive), the signal band $\mathcal{B}$ was defined over the range 0--0.5
Hz. In BGFixed, the analysis bands $\mathcal{A}_i$ was centered at a
frequency of interest $f_{c_i}$ with a fixed bandwidth of 0.1 Hz, for $0 \le
i \le I-1$. In contrast, BGAdaptive dynamically adjusted both the number of
tapers $K$ and analysis bandwidth $f_w$. Initially, each analysis band was
analyzed with four tapers and a bandwidth of 0.05 Hz. The number of tapers
was then iteratively increased until the maximum side lobe
leakage\footnote{For BGAdaptive, the \emph{side lobe leakage} of taper $k$
is calculated as $\gamma_k=10 \log_{10} \left(1-\lambda_k\right)$, where $1
\le k \le \gls{numtaper}$, which is given in TABLE I of Bronez,
1988~\cite{BRO1988}.} of the tapers was less than -10 dB. If this criterion
was not met after reaching the maximum of 8 tapers, the analysis bandwidth
was incremented by 0.01 Hz, and the process repeated. Iteration continued
until the leakage threshold was satisfied, or the bandwidth reached 0.5 Hz.
The final number of tapers and analysis bandwidth were then used to estimate
the power spectrum at the current frequency center. This adaptive process
was repeated for all frequency centers.

% note: figure width for one-column = 5in; two-column = 3.45in
\begin{figure}[!t]
	\centering
	\includegraphics[width=5.5in]{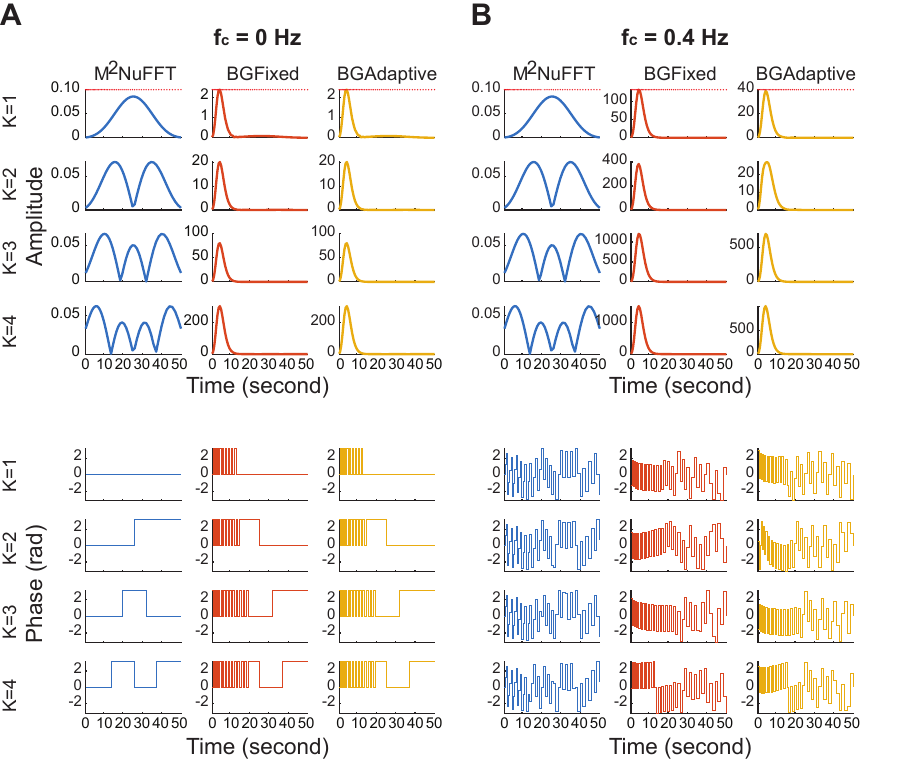}
	\caption{\textbf{Examples of Taper Amplitude and Phase for Arithmetic Sampling.}
		(\textbf{A}) Amplitude and phase of the first four tapers of the
		analysis band centered at $f_c = 0$ Hz, with a half-bandwidth of
		$0.05$ Hz.  The upper panel displays the amplitude, and the lower
		panel shows the phase.  In each panel, rows correspond to a
		different taper, while columns represent the methods: \algoname
		(interpolated DPSS), BGFixed, and BGAdaptive.  Red dots at the tops
		of each column indicate an instance of arithmetic sampling.
		(\textbf{B}) As in (\textbf{A}), but for an analysis band centered
		at $f_c = 0.4$ Hz.  Abbreviation: \textbf{\algoname},
		multiband-multitaper nonuniform fast Fourier transform;
		\textbf{BGFixed}, Bronez GPSS method with fixed TW~\cite{BRO1985};\
		\textbf{BGAdaptive}, adaptive Bronze GPSS method~\cite{BRO1985};
		\textbf{DPSS}, discrete prolate spheroidal
		sequence~\cite{THO1982}.}%
	\label{fig:taper_time_example}
\end{figure}

Figure~\ref{fig:taper_time_example} illustrates example tapers in the time
domain for the arithmetic sampling scheme, which exhibits the most
pronounced discrepancies among the four sampling scenarios considered (see
discussion below).  The figure shows the amplitude and phase of the first
four tapers at two analysis bands centered at $f_c = 0$ Hz and $f_c = 0.4$
Hz. At $f_c = 0$ Hz, \algoname\ tapers were approximated by interpolating the
corresponding DPSS.\@  At $f_c = 0.4$ Hz, the taper amplitude in \algoname\
remained identity to those at $f_c = 0$ Hz, except for a phase modulation
introduced by the NUFFT-based frequency shift. Note the higher sampling
density at the beginning of the arithmetic sampling instance (red dots at
the top of each column).  In contrast, Bronez GPSS methods (BGFixed and
BGAdaptive) adaptively adjusted the taper amplitude to local sampling
variability, resulting in taper shapes that differ from those of \algoname.
Although the Bronez methods produced similar amplitude profile across the
two analysis bands, the amplitudes scales differ besides phase modulation.

To quantify the deviation of \algoname\ tapers from the optimal solution, we
computed the root-mean-square error (RMSE) between the power spectra of
\algoname\ tapers and those of GPSS within $\pm f_w$ around each frequency
centers $f_{c_i}$, for $i = 0, 1, \ldots, I-1$, of the analysis bands
$\mathcal{A}_i$.  We further assessed discrepancies using the eigenvalue
differences defined in equation~\eqref{eq:subopt_measure}.

% note: figure width for one-column = 5in; two-column = 3.45in
\begin{figure}[!t]
	\centering
	\includegraphics[width=5in]{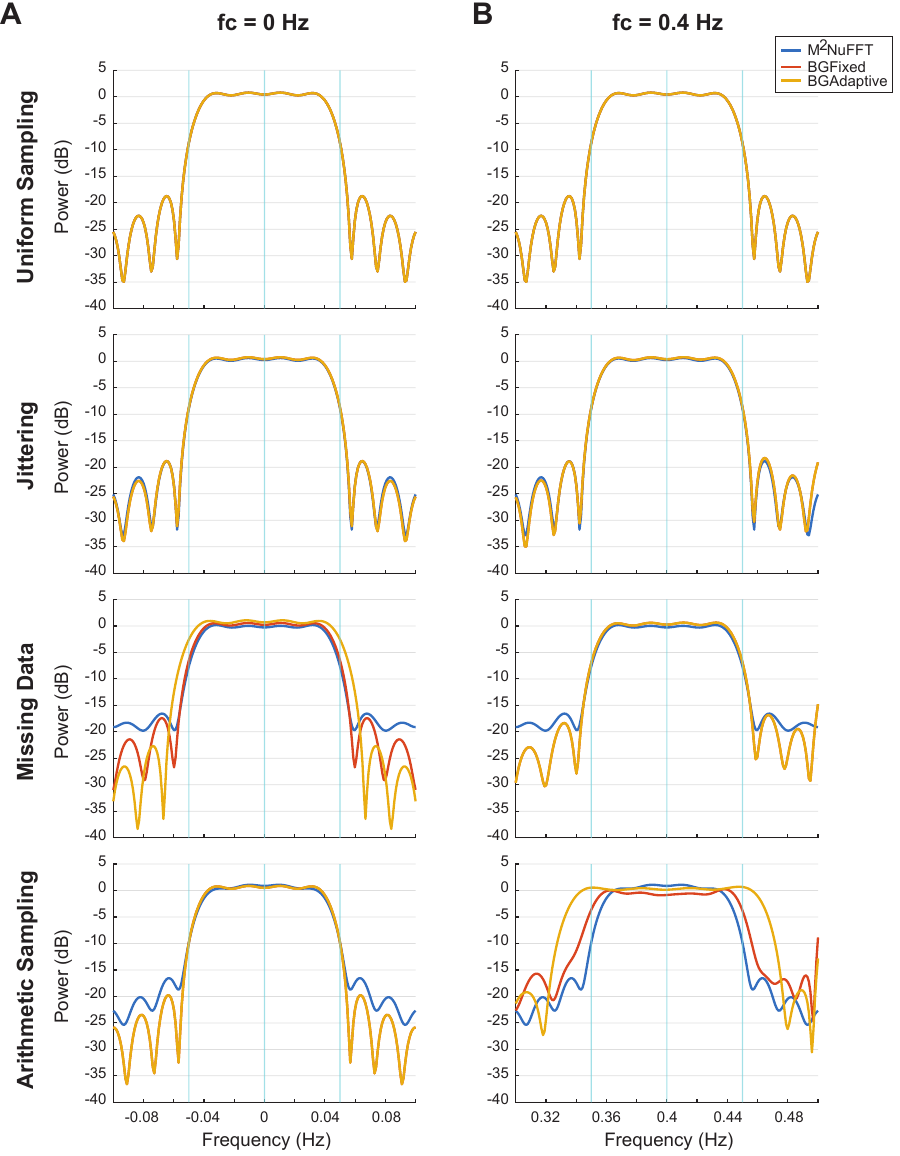}
	\caption{\textbf{Examples of Taper Power Spectra Across Sampling Methods.}
		(\textbf{A}) Average power spectra for four sampling
		schemes---Uniform sampling, Jittering, Missing data and Arithmetic
		sampling---estimated using three power spectrum estimation methods:
		\algoname, BGFixed and BGAdaptive at the analysis band centered at
		$f_c = 0$ Hz with a half-bandwidth of 0.05 Hz. The three cyan
		vertical lines indicate the center and the bound of the analysis
		band.  (\textbf{B}) The same as in (\textbf{A}) except $f_c = 0.4$
		Hz. Abbreviation: \textbf{\algoname}, multiband-multitaper
		nonuniform fast Fourier transform; \textbf{BGFixed}, Bronez GPSS
		method with fixed TW~\cite{BRO1985};\ \textbf{BGAdaptive}, adaptive
		Bronze GPSS method~\cite{BRO1985}; \textbf{DPSS}, discrete prolate
		spheroidal sequence~\cite{THO1982}.}%
	\label{fig:taper_freq_domain}
\end{figure}

Figure~\ref{fig:taper_freq_domain} illustrates the average taper power spectra
(computed as the DFT of the taper sequence; see
equation~\eqref{eq:taper_dft}) for the four sampling schemes and three
spectral estimation methods at the analysis bands centered at 0 Hz and 0.4
Hz.  As expected, under uniform sampling (top row), the taper spectra are
identical across methods.  In contrast, arithmetic sampling (bottom row)
exhibits substantial differences (c.f. Figure~\ref{fig:taper_time_example}),
reflecting the impact of nonuniform sampling on taper design.

% note: figure width for one-column = 5in; two-column = 3.45in
\begin{figure}[!t]
    \centering
    \includegraphics[width=5in]{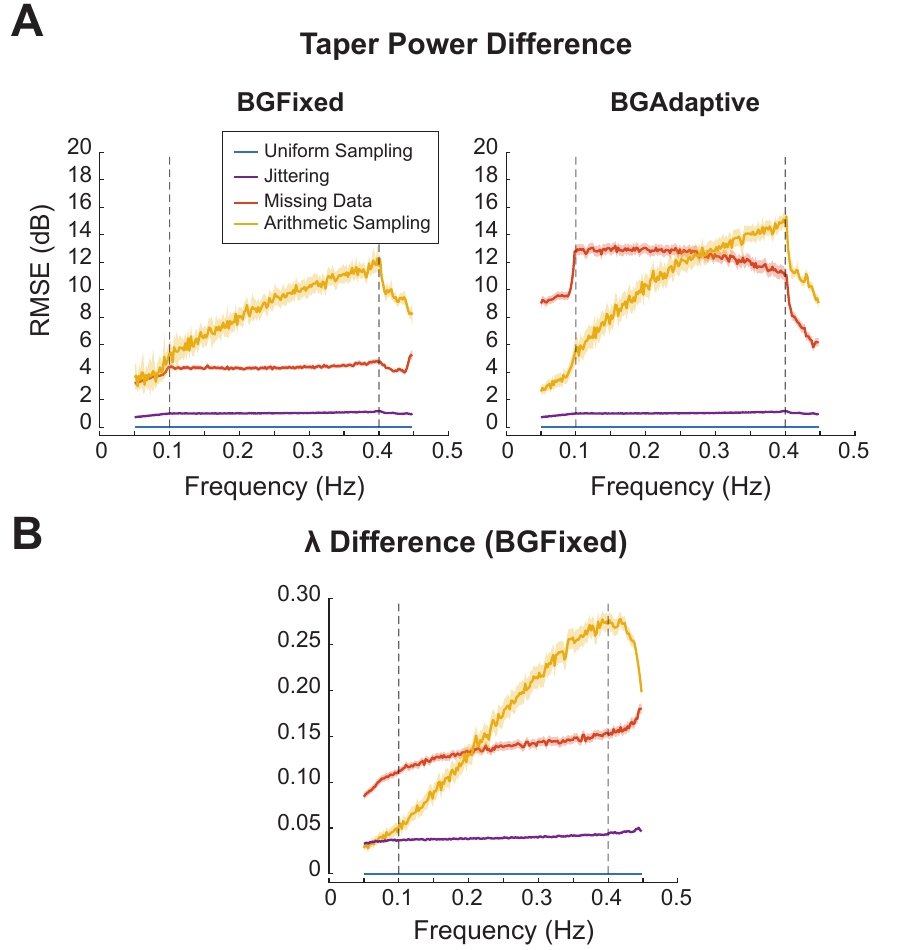}
    \caption{\textbf{\algoname\ Tapper Error Analysis.} (\textbf{A})
        \textit{Taper power difference}.  RMSE of the power difference
        between tapers estimated by \algoname\ and Bronez GPSS
        method---BGFixed (left panel) and BGAdaptive (right panel)---across
        four sampling schemes: Uniform sampling, Jittering, Missing data and
        Arithmetic sampling. Dashed vertical lines indicate the centers of
        the analysis bands at 0.1 Hz and 0.4 Hz, which are located $2f_w$
        away from the edges of the signal band. Estimated errors outside the
        dashed lines may be unreliable due to the boundary effect.
        (\textbf{B}) \textit{$\lambda$ difference [see
        Eq.~\eqref{eq:subopt_measure}]}.  Estimated average difference
        between the eigenvalues of the optimal GPSS tapers at the analysis
        band centered at 0 Hz and those centered at $f_{c_i}$, using BGFixed
        method. Dashed vertical lines are as in (\textbf{A}). Shaded error
        band: $\pm1$ standard error of the mean (SEM). Abbreviation:
        \textbf{RMSE}, root-mean-square error; \textbf{\algoname},
        multiband-multitaper nonuniform fast Fourier transform;
        \textbf{BGFixed}, Bronez GPSS method with fixed TW~\cite{BRO1985};\
        \textbf{BGAdaptive}, adaptive Bronze GPSS method~\cite{BRO1985};
        \textbf{GPSS}, generalized prolate spheroidal
        sequence~\cite{THO1982}.}%
    \label{fig:taper_err_analysis}
\end{figure}

We summarize the taper error analysis in
Figure~\ref{fig:taper_err_analysis}. Panel (\textbf{A}) shows taper power
difference. For uniform sampling, no differences were observed since all
methods produced identical tapers. For jittering and missing-data sampling,
the errors remain relatively constant across the signal band, with
missing-data sampling inducing significantly larger errors than jittering
method. Interestingly, the arithmetic sampling errors increase approximately
linearly with frequency, suggesting that the nearly symmetric tapers of
\algoname\ fail to capture local variations in sampling density at higher
frequencies. Moreover, BGAdaptive generally exhibits larger errors than
BGFixed, likely due to its lower sidelobe leakage achieved through adaptive
procedure.

Panel (\textbf{B}) presents the eigenvalue difference analysis for BGFixed.
The pattern mirrors that of Panel \textbf{A}---relatively constant
differences for jittering and missing data sampling, and a linear increase
for arithmetic sampling. This consistency supports the usage of
equation~\eqref{eq:subopt_measure} as a valid suboptimality metric.  Note
that the eigenvalue differences for BGAdaptive are not reported because the
number of tapers varies cross analysis bands in the adaptive approach.

It is worth noting that the taper power difference analysis relied on
interpolated DPSS to approximate the optimal tapers at the nominal band
centered at 0 Hz, whereas the $\lambda$-difference analysis used the true
optimal tapers at the same band.  The strong similarity between these two
sets of results suggests that the error introduced by DPSS interpolation is
likely insignificant, at least for the sampling schemes considered in this
study.

\section{Performance Evaluation}\label{sec:per_eval}
Due to the complexity of the proposed \algoname\ algorithm, in this section
we only consider the performance on the full signal band, i.e., $B = \{f:
|f| \le \gls{fmax}\}$, and the nominal band \gls{A0} centered at zero
frequency, which is usually the case in practice.  We evaluated the
performance in three key aspects: accuracy, speed, and real-world
applicability. Initially, we computed the Mean-Square Error (MSE) between
the estimated and the actual power spectra of Gaussian white noise (GWN)
under various sampling schemes. The results indicated that the error range
of \algoname\ was compatible with that of the optimal method, BGAdaptive.
Subsequently, we contrasted the speed of \algoname\ with three alternative
methods. Our findings revealed that the speed of our algorithm is 2--3
orders of magnitude higher than that of the optimal method. Lastly, we
applied our method to estimate the power spectrum of a real-world signal,
specifically a nonuniformly sampled impedance measurement. We then compare
the outcomes of Thomson's F-test on the periodicity of both the original and
resampled signals. This comparison allows us to evaluate the effectiveness
of our method in practical applications.

% note: figure width for one-column = 5in; two-column = 3.45in
\begin{figure}[!t]
    \centering
    \includegraphics[width=4.5in]{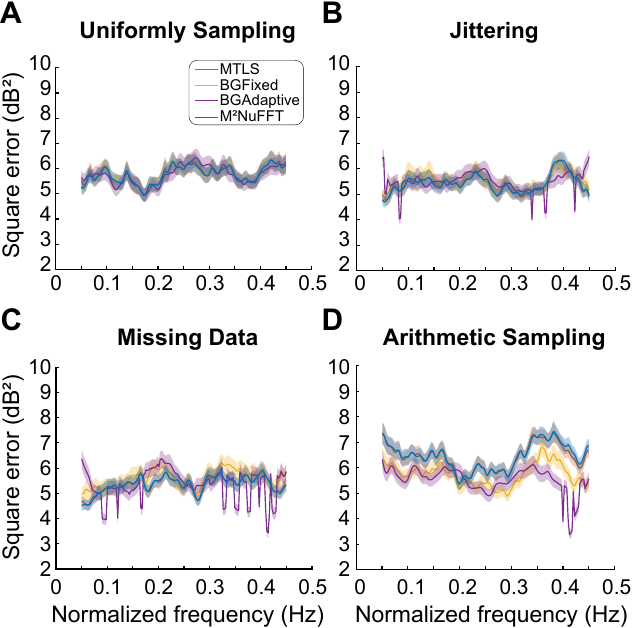}
    \caption{\textbf{Error Analysis of Spectrum Estimation Methods.}
        Mean-square error comparison between the true spectrum (Gaussian
        white noise, GWN) and the estimated spectra using four estimation
        approaches, i.e., MTLS, BGFixed, BGAdaptive, and \algoname, and the
        signals were sampled using four sampling schemes: (\textbf{A})
        Uniformly Sampled, (\textbf{B}) Jittering, (\textbf{C}) Missing
        Data, and (\textbf{D}) Arithmetic Sampling. The frequency range was
        normalized to 0--0.5 Hz and half-band width ($f_w$) was set at 0.05
        Hz. Error measures at 0--0.5 and 0.45--0.5 Hz were omitted due to
        unreliable estimation. Error band: $\pm 1$ SEM. Abbreviation:
        \textbf{MTLS}, multitaper Lomb-Scargle periodogram~\cite{SPR2020};
        \textbf{BGFixed}, Bronez GPSS method with fixed TW~\cite{BRO1985};
        \textbf{BGAdaptive}, adaptive Bronez GPSS method~\cite{BRO1985};
        \textbf{\algoname}, multiband-multitaper nonuniform fast Fourier
        transform (* proposed in this article).}%
    \label{fig:gwn_error_analysis}
\end{figure}

\subsection{Error Analysis}%

We assessed the accuracy of the proposed fast algorithm by comparing the
estimated spectrum with the true spectrum of unit variance GWN ($\sigma^2 =
1$), which was conducted using four estimation methods (MTLS, BGFixed,
BGAdaptive, and \algoname) and the four sampling schemes described above
(Section~\ref{sec:taper_subopt}) to sample 50 points from the GWN.\@ 

The multitaper methods adopted in \algoname, BGFixed and BGAdaptive were
described in Section~\ref{sec:taper_subopt}.  The parameters of the tapers for
MTL were identical to those of \algoname.

We repeated the process to evaluate the power spectrum of the GWN $M = 1000$
times for each estimation method and each sampling scheme. Subsequently, we
computed the MSE in decibels (dB) between the estimated spectrum
$\hat{S}(f_{c_i})$ and true spectrum $S(f_{c_i})$ at each frequency center
$f_{c_i}$, which was calculated as $\text{MSE}(f_{c_i})=\frac{1}{M} \sum_{m
    = 1}^M {[10\log_{10}\hat{S}(f_{c_i})]}^2$, given that $S(f_{c_i}) = 1$.

Figure~\ref{fig:gwn_error_analysis} presents the error analysis results,
organized into four panels that corresponds to the four sampling schemes.
Each panel displays the mean and standard error of the mean (SEM) of squared
error at each frequency center. For uniformly sampled signal, the error
range was essentially identical across all four estimation methods. The
\algoname\ method demonstrated a compatible error range to the Bronez GPSS
methods (BGFixed, BGAdaptive) when applied to jittering and missing-data
sampling. However, the BGAdaptive method exhibited a better performance at
isolated frequency centers. In the case of the arithmetic sampling scheme,
the Bronez GPSS methods marginally yet significantly outperformed both MTLS
and \algoname\ across the entire signal band $\mathcal{B}$. Overall, the
proposed fast algorithm \algoname\ demonstrated competitive accuracy in
spectrum estimation in three of the four sampling schemes investigated.

\subsection{Speed Analysis}
\begin{figure}[!t]
    \centering
    \includegraphics[width=4.5in]{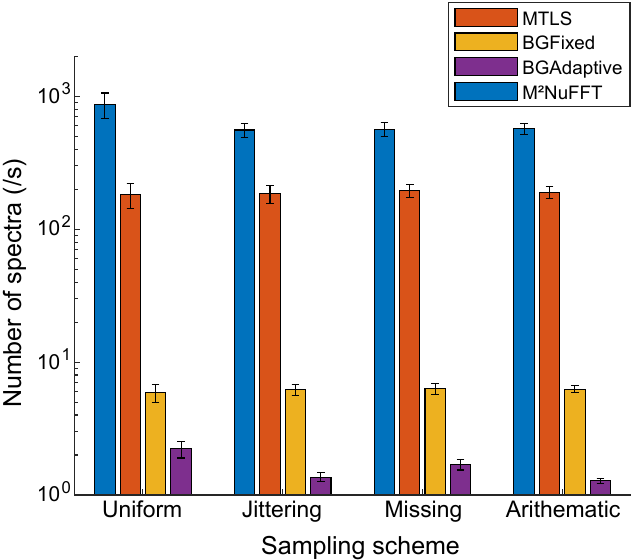}
    \caption{\textbf{Speed Analysis of Spectrum Estimation Methods.} This
        figure presents the number of spectra calculated per second for four
        sampling schemes using four spectrum estimation methods. The
        abbreviations used are consistent with those in
        Figure~\ref{fig:gwn_error_analysis}. Error bar: $\pm 1$ STD}%
    \label{fig:speed_analysis}
\end{figure}

The time efficiency of the proposed method was assessed by comparing the
number of spectra computed per second across four different sampling
schemes, using the four spectrum estimation methods. The performance
evaluation was conducted on a Windows 10 HP workstation equipped with an
Intel Core i5--10500 CPU operating at 3.10 GHz and 64 GB memory. We
estimated the time cost for 1,000 spectrum estimation and obtained the mean
and standard deviation (STD). The results, as presented in
Figure~\ref{fig:speed_analysis}, indicate that the speed of \algoname\
significantly surpasses that of MTLS and 2--3 orders of magnitude faster
than the Bronez GPSS approaches for all four sampling schemes investigated.

% one-column width = 5in; two-column width = 3.45in

\subsection{Application to Impedance Signal}
\begin{figure}[!t]
    \centering
    \includegraphics[width=5.5in]{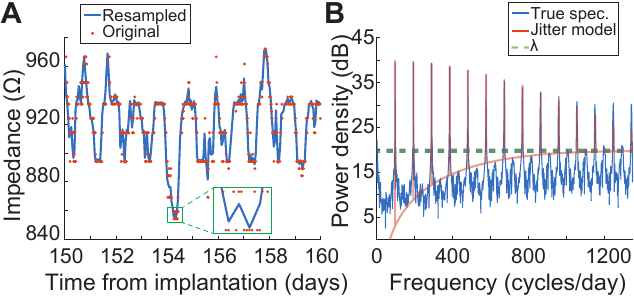}
    \caption{\textbf{Impedance Signal Sampling and Jitter Model Fitting.}\
        (\textbf{A}) An example of impedance signal nonuniformly sampled at
        approximately four samples per hour using a Medtronic Summit
        RC+S\texttrademark\ neuromodulation device. Red dots represent
        original samples obtained 150--160 days post-implantation. The blue
        curve shows the resampled signal at a uniform rate of one sample per
        hour.  The green rectangular inlet is a zoomed signal interval,
        indicating highly packed impedance samples (red dots). (\textbf{B})
        Power spectrum of the sampling points from (A) and its fitted model.
        The blue curve represents the sampling process spectrum (calculated
        with \textsc{Chronux} function \texttt{mtspectrumpt}, $TW = 3.5$ and
        $K = 6$), while the red curve depicts the fitted model, a uniformly
        sampled point process with jitter (see text for more details). The
        green dashed line marks the average sampling rate ($\lambda = 96$
        /day or $10\log_{10}(96) = 19.82$ dB). Abbreviation: \textbf{spec},
        spectrum.}%
    \label{fig:example_and_jitter_model_fit}
\end{figure}

To further illustrate the \algoname\ method, we applied it to the spectral
analysis of a bio-impedance signal recorded intracranially from human brain
using a chronically implanted sensing and stimulation device. The specifics
of the brain impedance acquisition and analysis have been detailed in our
prior work~\cite{MIV2023,CUI2024}. The impedance signal was measured using
the investigational Medtronic Summit RC+S\texttrademark\ device with the
electrodes targeting the limbic system of a patient with epilepsy. Given
that the same electrodes were also utilized for delivering electric
stimulation as part of neuromodulation therapy, the impedance measurements
were nonuniformly sampled at an approximate rate of one sample every 15
minutes, equivalent to about 96 samples/day.
Figure~\ref{fig:example_and_jitter_model_fit}A shows a data segment of
between 150 and 160 days post device implantation (number of sample, $N =
688$), which was used in the analysis. The red dots represent the original
impedance samples, while the blue curve signifies the resampled signal at a
uniform rate of one sample per hour ($N = 240$, calculated with
\textsc{Matlab} function \texttt{resample} using linear interpolation). In
Panel B, the blue curve illustrates the power spectrum of the
\textit{sampling instances} of the original impedance signal. The decaying
envelope of the sharp lines at the fundamental frequency of 96 cycles/day
and its harmonics are indicative of irregular
sampling~\cite{BAR1963,JAR2001,BAL1962}. We fitted the spectrum with a
jittering model~\cite{BRE2014}, assuming a normal distribution of the
jittering displacement $z_n$ with zero mean and STD $\sigma$. The red curve
in Figure~\ref{fig:example_and_jitter_model_fit}B represents the fitted
model with mean rate $\lambda = 96$ samples/day and STD $\sigma = 20$
seconds, and the green dashed line indicates the mean rate at high-frequency
limit. This model offers a good understanding of sampling properties of the
impedance measurement sequence.

% one-column width = 5in; two-column width = 3.45in

Subsequently, we computed the power spectra of the original signal and the
resampled signal using \algoname\ and \textsc(Chronux) \texttt{mtspectrumc},
respectively, under the assumption of a maximum frequency of 12 cycles/day.
Identical to the calculation of point process power spectrum, the TW was set
at 3.5, yielding a frequency resolution of 0.35 cycles/day, and $K$ at 6. As
displayed in Figure~\ref{fig:spectrum_and_f_test}A, the spectral power of
the nonuniformly sampled signal (represented by the red curve) is noticeably
elevated above approximately 2 cycles/day in comparison to the spectral
power of the resampled signal (blue curve). This observation aligns with
previous studies~\cite{SPR2020,LEP2009} (see\
Section~\ref{sec:introduction}: Introduction). While the spectral powers are
nearly identical around the circadian cycle (1 cycle/day, indicated by arrow
a) and multiday cycles ($<$ 1 cycle/day), distinct energy peaks are
presented in the frequency ranges of 2--4 Hz (arrow b) and 4--6 Hz (arrow
c), which are absent in the power spectrum of the resampled data. However,
the elevated power at high-frequency range of 10--12 Hz (arrow d) could
potentially be attributed to leakage (refer to
Section~\ref{sec:discussion}~Discussion).

\begin{center}
    \begin{figure}[!t]
        \centering
        \includegraphics[width=6.0in]{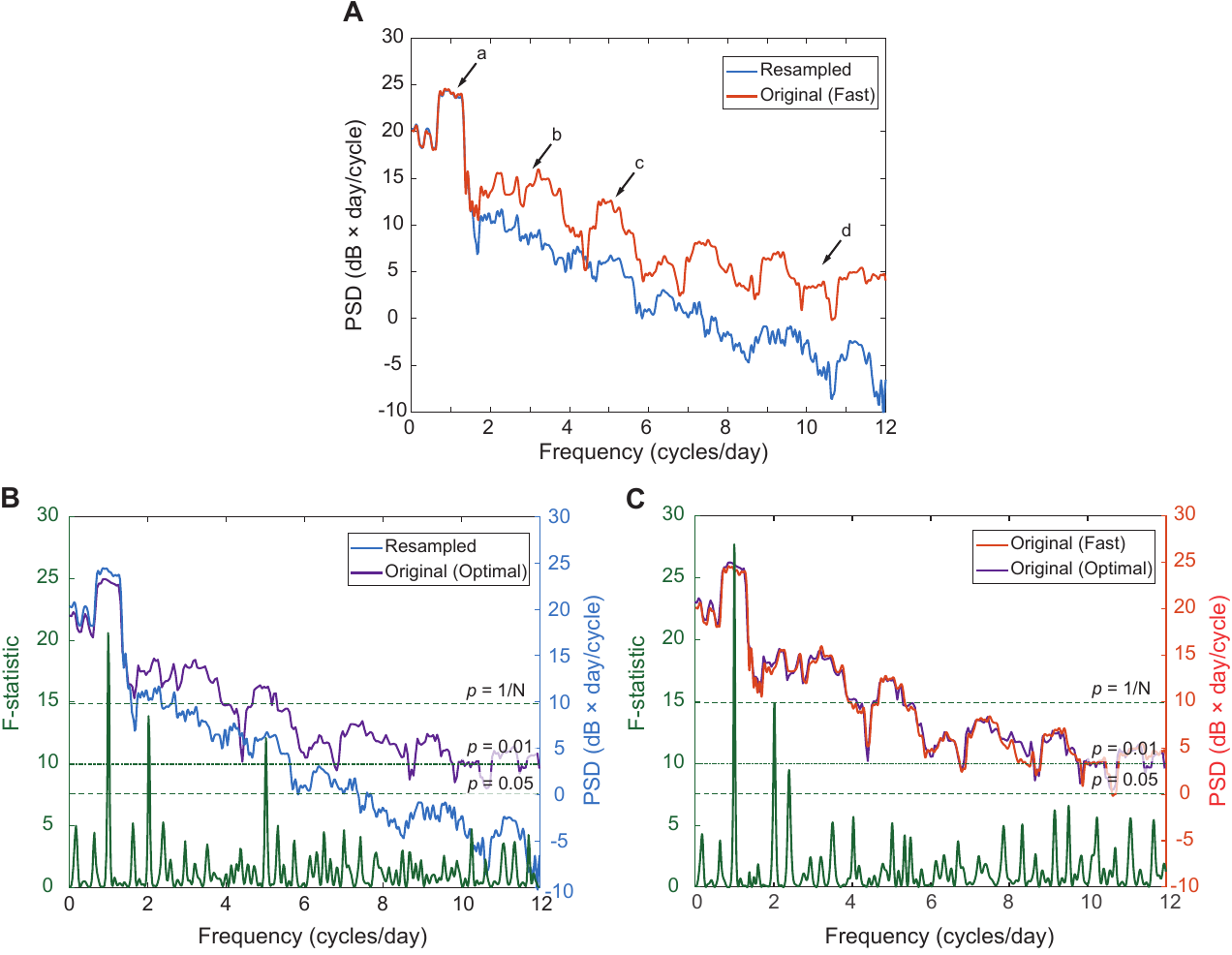}
    \end{figure}
    \captionof{figure}[F-test]{\textbf{Spectrum and F-test Comparison.}\
    (\textbf{A}) Spectrum comparison. The red curve (Original (Fast))
    represents the estimated power spectrum of the nonuniformly sampled
    impedance data (referenced as red dots in
    Fig.~\ref{fig:example_and_jitter_model_fit}A) using MTNUFFT .\ The blue
    curve (Resampled) depicts the power spectrum of the resampled signal
    (blue curve in Fig.~\ref{fig:example_and_jitter_model_fit}A), estimated
    with the \textsc{Chronux}~\cite{BOK2010} function \texttt{mtspectrumc}.
    The arrows point to approximate frequency bands of circadian cycles (a,
    1 cycle/day), 2--4 (b), 4--6 (c), and 10--12 cycles/day (d).
    (\textbf{B}) Power spectrum of the resampled signal (blue, same as in
    Panel A), power spectrum of the original signal estimated with the
    optimal method \texttt{BGFixed} (purple, original (Optimal)), and
    Thomson periodicity F-test~\cite{THO1982} (green) of the resampled
    signal.\ (\textbf{C}) Power spectrum estimated with the fast method
    (red, same as in Panel A, original (Fast)), power spectrum estimated
    with the optimal method (purple, same as in Panel B, Original
    (Optimal)), and F-test (green) of the original, nonuniformly sampled
    signal. In Panels B and C, three horizontal dashed lines represent three
    levels of \textit{p}-values, i.e., from bottom to top, 0.05, 0.01 and
    $1/N$, respectively, where $N$ is the number of samples. For the
    resampled signal (Panel B), $N = 240$, resulting in $p = 0.0042$. For
    the original signal (Panel C), $N = 688$, yielding $p = 0.0015$.}%
    \label{fig:spectrum_and_f_test}
\end{center}

Moreover, we investigated the periodicity of impedance data using the
Thomson \textit{F}-tests~\cite{THO1982,PER1993}. Specifically, we analyzed
both the original (Figure~\ref{fig:spectrum_and_f_test}B) and resampled data
(Figure~\ref{fig:spectrum_and_f_test}C), where the optimal estimation of the
spectrum using BGFixed is superimposed for comparison. To assess
significance, we calculated critical values of the \textit{F}-statistic
corresponding to three significant levels: \textit{p}-values at 0.05, 0.01
and the Rayleigh level ($1/N$, where $N$ is the number of samples). These
critical values were derived from the \textit{F}-distribution with 2 and
$2K-2 = 10$ degrees of freedom. It is worth noting that the Rayleigh level
($p = 1/N$), as recommended in Thomson et al.~\cite{THO1982}, bears
similarity to the Bonferroni correction for multiple
comparison~\cite{CHR2011,BRE2007}. The analysis reveals some intriguing
findings. The \textit{F}-statistic for the resampled signal (Panel B)
indicates a strong periodicity in the circadian cycle (above the Rayleigh
level) and suggests two possible cycles around 2 and 5 cycles/day (above $p
= 0.01$ level). In contrast, the \textit{F}-statistic for the original
signal confirms the robust periodicity of the circadian cycle and its
harmonic at 2 cycles/day (above the Rayleigh level). Notably, it also
depicts the absence of the periodicity around 5 cycles/day, raising the
possibility that this suggested periodicity in the resampled signal may
result from linear interpolation.

\begin{figure}[!t]
    \centering
    \includegraphics[width=4.5in]{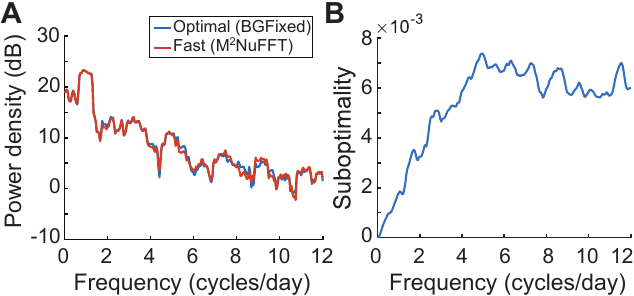}
    \caption{\textbf{Spectrum comparison and suboptimality.}\ (\textbf{A})
        The optimal spectrum estimated with BGFixed method (blue) is
        compared with the spectrum estimated with the fast algorithm (red,
        \gls{m2nufft}), where the eigenvectors were $\gls{wk0}, \quad 1 \le
        k \le \gls{numtaper}$ (\ref{eq:gep_at_A0}), optimal at
        $\mathcal{A}_0$. (\textbf{B}) Suboptimality
        (\ref{eq:subopt_measure}) of the spectral estimation \gls{m2nufft}
        shown in Panel A.}%
    \label{fig:fast_algorithm_comparison}
\end{figure}

These analyses of the ultradian cycles (occurring more frequently than once
per day) within impedance signals hold significant biological interest.
Long-standing hypotheses propose that an ultradian basic rest-activity cycle
(BARC) plays a crucial role in sleep cycles, wakefulness patterns, and the
central nervous system functioning~\cite{KLE1982}.

Finally, we evaluated the suboptimality of the power spectrum estimation of
the impedance signal by calculating \gls{E}~(\ref{eq:subopt_measure}). Given
that the approximate eigenvectors \gls{wk0est} lack corresponding
eigenvalues, we utilized the true eigenvectors \gls{wk0} at \gls{A0} for the
fast algorithm \gls{m2nufft} (see Table \ref{tb:m2nufft_method}), denoted as
M\textsuperscript{2}NuFFT0. The power spectra, estimated by both the optimal
method (BGFixed) and the fast algorithm (\algoname0), are depicted in
Figure~\ref{fig:fast_algorithm_comparison}A. It's important to note the
identity of the spectra at $f=0$,  which is confirmed by the suboptimality
measure $\mathcal{E} = 0$ at $f = 0$, as shown in
Figure~\ref{fig:fast_algorithm_comparison}B.  As previously discussed,
suboptimality $\mathcal{E}$ is between 0 and 1, where zero signifies
optimality. The larger this measure, the greater the deviation of the
estimation is from the optimal scenario.  The suboptimality increases as the
center of the analysis band shifts away from $f = 0$, but appears to plateau
after about $f = 5$ cycles/day. Overall, the suboptimality is less than $8
\times 10^{-3}$, suggesting that the proposed method effectively estimated
the power spectrum of the impedance signal.

\section{Discussion}\label{sec:discussion}
We have developed \gls{m2nufft}, a method for a rapid and scalable power
spectrum estimation of nonuniformly sampled time series. This method
alleviates the computational burden of the Bronez \gls{gpss} in three key
aspects: (\textbf{1}) The multiband framework enables partitioning of the
signal band into sub-bands, which is well suited for parallel computing
architecture, allowing substantial computational time reduction.
(\textbf{2}) The algorithm requires solving the computationally intensive
\gls{gep} only once per sub-band.  The resulting optimal tapers \gls{wk0}
for \gls{A0} are then shifted to other analysis bands \gls{Ai}, for $1 \le i
\le \gls{numAi}-1$, using the \gls{nufft} to circumvent the remaining
\gls{gep} problems. (\textbf{3}) In the special case when \gls{A0} is
centered at zero frequency, the optimal taper \gls{wk0} can be efficiently
approximated by interpolating the \gls{dpss} tapers, \gls{dpsskn}, to the
nonuniform grid using cubic splines~\cite{SPR2020}. This eliminates the need
for \gls{gep} entirely. As a result, the overall computational cost aligns
with the fast \gls{nufft}, $O(N \log N + N \log(1/\epsilon))$, which is
significantly lower than the $O(N^4)$ complexity of the optimal method.
Simulations (Figure~\ref{fig:speed_analysis}) show that the computing speed
of \gls{m2nufft} is significantly higher than \gls{mtls}, and 2--3 orders of
magnitude faster than the \gls{bg} methods (both BGFixed and BGAdaptive) in
the zero-centered case.

Bronez~\cite{BRO1985,BRO1988} also proposed a computationally efficient
approximation known as the Constrained-Basis Weighting Sequences method. The
basic idea is to reduce the size of the matrix in the GEP (\ref{eq:gep})
from $N$ to $M$, ideally $M \ll N$. This is achieved by selecting $M$
``basis vectors'' and approximating the weight sequence as
$\widehat{\mathbf{w}}_k(\mathcal{A}) = F \cdot \mathbf{c}_k(\mathcal{A})$,
where $F$ is an $N \times M$ matrix with columns as the predefined basic
vectors, and $\mathbf{c}_k(\mathcal{A})$ are $M \times 1$ vectors determined
by solving another GEP for analysis band $\mathcal{A}$. However, the matrix
size of this problem is only $M$. The vector $\mathbf{c}_k(\mathcal{A})$
still needs to be calculated for each analysis band. The overall computing
load is in the order of $O(M^2 N^2)$, which can be significantly lower than
$O(N^4)$, but still notably higher than that of \gls{m2nufft}. The choice of
the basis vectors is critical to the method's performance and requires
careful consideration.

Since \gls{m2nufft} is not an optimal solution, it is crucial to evaluate the
deviation from the optimal solution of its estimates. Our theoretical work
(Section~\ref{sec:m2nufft}) and numerical experiments
(Section~\ref{sec:taper_subopt}) show that the bias of estimation and
variance bound are compatible with the optimal Bronez GPSS.\@ But the bias
bound is generally degraded~\cite{BEL1997}. The suboptimality of \gls{m2nufft}
for each analysis band $\mathcal{A}_i$ may be quantified by the difference
between $\lambda_k^0$ and $\lambda_k^i$ (\ref{eq:subopt_measure}), which
decreases at the expense of increasing analysis band (decrease of frequency
resolution). The simulation results (Figure~\ref{fig:gwn_error_analysis})
show that for the four sampling schemes under investigation, the error range
is compatible with the optimal method for uniformly sampling, jittering and
missing-data sampling. Only the error range for arithmetic sampling is
consistently higher than the optimal methods. These results indicate the
effectiveness of the proposed method for most practical scenarios. Although
our analysis attributes the suboptimality less to the interpolation of DPSS
than to the frequency shifting of tapers via \gls{nufft}, the likely loss of
orthogonality of the interpolated DPSS remains an interesting topic for
future research~\cite{GEO2025}.

It is worth noting that the variance and bias bounds may not be accurately
estimated when the analysis bands are near the boundary of signal band,
$f_{\max}$. In deed, a previous study~\cite{WAL1994} showed that the
variance of spectrum estimates could be poorly estimated if the frequency
was close to frequency limits.

As noted in Section~\ref{sec:m2nufft}, the algorithm proposed by Patil et
al.~\cite{PAT2024a,PAT2024b}, referred to as mtNUFFT, corresponds to a
special case under our more generalized \gls{m2nufft} framework. Their work
provided compelling empirical evidence for the practical effectiveness of
their algorithm, particularly for astrophysical data analysis. However, our
contribution extends beyond empirical validation by offering a theoretical
framework that quantifies the estimator's performance in terms of bias,
variance and suboptimality, which were not sufficiently addressed in prior
work.  This theoretical development allows us to recognize that mtNUFFT and
Bronze \gls{gpss} method represent two extremes under the same \gls{m2nufft}
framework. At one end, mtNUFFT assumes a single signal band centered at zero
frequency with interpolated DPSS tapers; at the other, Bronez's method
computes frequency-dependent tapers for each analysis band. Our framework
bridges these approaches, offering a continuum of trade-offs between
computational efficiency and statistical optimality. 

Moreover, our normalization approach, based on
equation~\eqref{eq:bg_norm_requirement}, ensures signal energy conservation
(c.f. Figure~\ref{fig:taper_freq_domain}). This insight is not readily
apparent through numerical evaluation alone.  In contrast, the $L^2$-norm
normalization commonly used for uniformly sampled \gls{dpss} may introduce a
constant bias in the power spectrum estimate. While this bias my be
negligible in practice, it underscores the importance of theoretically
grounded normalization. Patil et al.\ also emphasize the importance of
quasi-regular time sampling for the effectiveness of mtNUFFT.\ Our analysis
provides a theoretical rationale for this observation. As seen in
Section~\ref{sec:taper_subopt} and~\ref{sec:per_eval}, jittered and
missing-data sampling tend to produce a relatively constant and smaller
deviation from the optimal solution (Figures~\ref{fig:taper_err_analysis}
and~\ref{fig:gwn_error_analysis}). These sampling patterns preserve the
structure of the signal sufficiently well to maintain taper performance,
making them favorable for suboptimal methods like \gls{m2nufft}.

The process of resampling nonuniformly sampled values to a uniform grid, for
example, using linear interpolation, is often employed for spectral analysis
due to the powerful tools available for estimators under uniform sampling.
However, evidence suggests that this procedure could induce considerable
artifacts in the power spectrum~\cite{LEP2009}. One noticeable effect of
linear interpolation is a tendency for the estimated spectrum at high
frequency to be lower, and at low frequency to be higher~\cite{SPR2020}.
This distortion can be substantial~\cite{MAN1996}. Our analysis of the brain
tissue impedance data also indicates a suppression of spectral power at
higher frequencies due to interpolation. As shown in
Figure~\ref{fig:spectrum_and_f_test}A, the spectrum estimated from original
samples using \gls{m2nufft} (red curve) and the spectrum estimated from the
resampled signal (blue curve) are nearly identical at lower frequency ranges
($< 1.5$ cycles/day), but notably different at higher frequency ranges
(1.5--12 cycles/day). This observation is consistent with the previous
findings and underscores the need to develop spectral estimators that
directly utilize nonuniformly sampled data.

The \gls{m2nufft} algorithm may be considered as a general framework for
quickly estimating the spectrum of nonuniformly sampled signals using
various types of tapers. Besides DPSS sequences, other tapers, such as
minimum bias tapers and sinusoidal tapers~\cite{RIE1995,PER2020}, have been
previously suggested for different problems. Generally, these methods have
not been extended to the case of nonuniformly sampled signals. By replacing
the weight sequence $\mathbf{w}_k^0$ in (\ref{eq:eigencoefficients}) with
the desired tapers, which are properly evaluated on the nonuniform sampling
grid, and then applying the \gls{nufft} and averaging
(\ref{eq:power_spectrum_estimator}), the \gls{m2nufft} algorithm may be
extended to these tapers for spectrum estimation in nonuniformly sampled
time series. Future work will focus on extending \gls{m2nufft} to different
tapers in various data analysis scenarios and evaluating the statistical
properties of the estimation.

A limitation of this study is that the theoretical analysis of \gls{m2nufft}
optimality relies on minimizing the estimation bounds of Gaussian white
noise spectrum.  While this optimality criterion has proven effective for
evaluating the performance of the fast algorithm in neurophysiological
signal spectral estimation---our main targeted field of application---it may
not be universally suitable.  Specifically, an optimality criterion based on
colored noise could be more appropriate for other applications. Future work
will focus on extending the theoretical analysis to these scenarios.

The current \gls{m2nufft} algorithm assumes that the partitioning of
 sub-signal bands, \gls{subB}, and the bandwidth, \gls{halfbw} (determined
 by the number of tapers \gls{numtaper}), are pre-defined (see
 Table~\ref{tb:m2nufft_method}). However, the proposed framework suggests a
 potential pipeline for adaptive and iterative refinement of these
 parameters. As a first step, a cluster analysis of the sampling times could
 be performed to identify regions of relatively dense sampling, separated by
 large temporal gaps.  Within each cluster, an initial spectral estimate can
 be obtained using a fast implementation of \gls{m2nufft} (e.g.,
 $\gls{numsubB} = 1$ and $f_0 = 0$) based on an initial guess of sub-band
 structure and bandwidth. Subsequent refinement can then be guided by the
 characteristics of the preliminary spectrum: for regions where the spectrum
 is smooth and slowly varying, fewer and wider sub-bands (i.e., decrease
 \gls{numsubB} and increase \gls{halfbw}) may suffice; conversely, regions
 exhibiting rapid spectral changes may benefit from finer segmentation
 (i.e., increase \gls{numsubB} and decrease \gls{halfbw}).  This adaptive
 procedure can be repeated iteratively to improve spectral estimates within
 each cluster. For critical frequency regions---such as those associated
 with critical physiological or physical phenomena---the Bronez \gls{gpss}
 method can be selectively applied to obtain optimal spectral estimates for
 the corresponding sub-bands. Finally, the spectral estimates from all
 clusters can be aggregated to produce a comprehensive and refined power
 spectrum. A fully automated implementation of this iterative refinement
 strategy represents an interesting direction for future work.

\section{Summary}\label{sec:summary}
This paper introduces \gls{m2nufft}, a fast suboptimal method for power
spectrum estimation in nonuniformly sampled time series. The proposed
multiband-multitaper framework leverages parallel computing architecture and
avoids the computational bottlenecks of Bronez \gls{gpss} method. In each
sub-signal band, the estimator comprises a set of nonuniformly sampled
tapers, optimized for a nominal analysis band. The estimated power within
the band is determined by averaging the power correlated to these tapers.
The \gls{nufft} is utilized to swiftly shift the tapers to other analysis
bands of interest, thereby removing the need to solve the \gls{gep}
repeatedly.  In the special case where the nominal band is centered at zero
frequency, the \gls{gep} computation can be circumvented entirely by
approximating the \gls{gpss} tapers with the interpolated \gls{dpss} tapers.
The overall computational complexity of \gls{m2nufft} in this case aligns
with that of \gls{nufft}, $O(N \log N + N \log(1/\gls{preci}))$.

The statistical properties of the estimator are assessed using the Bronez
\gls{gpss} theory. The results reveal that the bias of the estimates and
variance bound of \gls{m2nufft} are comparable to those of the optimal
estimator. However, the limitation of \gls{m2nufft} lies in the degradation
of the bias bounds. The difference in bias bounds between \gls{m2nufft} and
the optimal estimator may serve as a measure of suboptimality. Simulation
results indicate that \gls{m2nufft} operates 2--3 orders faster than the
optimal method. Moreover, the error range of \gls{m2nufft} aligns with that
of the optimal estimator in three out of four sampling schemes under
investigation, suggesting effectiveness in practical applications. The
\gls{m2nufft} together with the proposed extension of Thomson
\textit{F}-test, is suitable for rapid spectrum estimation and periodicity
testing in large nonuniformly sampled datasets for exploratory analysis.

\addcontentsline{toc}{section}{Acknowledgment}\label{sec:acknowledgment}
\section*{Acknowledgment}
The authors would like to thank Dr.~Filip Mivalt, Vladimir Sladky, and
Dr.~Vaclav Kremen for providing the human brain bio-impedance data.  This
work is supported by US National Institutes of Health (NIH) grants
UH3-NS095495, R01-NS092882 and R01-NS112144 (to G.W.). J.C. was also
partially supported by Epilepsy Foundation of America's My Seizure Gauge
grant (to B.H.B), US NIH grant UG3-NS123066 (to B.H.B.), and Mayo Clinic RFA
CCaTS-CBD Pilot Awards for Team Science UL1TR000135 (to J.C.).

% **************
% appendix
% **************
\glsresetall%
% The Appendices part is started with the command \appendix; appendix
% sections are then done as normal sections
\appendix
\section{GPSS Matrix of Sub-Signal Band is Positive Definite Hermitian}\label{app:pos_def_herm}
According to the definition shown in~(\ref{eq:gpss_matrix_b}), the
\gls{gpss} matrix of sub-band \gls{subB} is given by
\begin{align*}
    \mathbf{R}(\gls{subB}) = \mathbf{R}(\mathcal{A}_1) - \mathbf{R}(\mathcal{A}_2),
    \numberthis%
\end{align*}
where $\mathcal{B}^q = \{f: f_{\min}^q \leq |f| \leq f_{\max}^q \}$,
$\mathcal{A}_1 = \{f: |f| \leq f_{\max}^q \}$, and $\mathcal{A}_2 = \{f: |f|
< f_{\min}^q \}$.  Note that $\gls{subB}=\gls{A}_1-\gls{A}_2$. Since both
$\mathbf{R}(\mathcal{A}_1)$ and $\mathbf{R}(\mathcal{A}_2)$ are Hermitian,
their difference $\mathbf{R}(\gls{subB})$ is also Hermitian.

To show that \gls{gmatB} is positive definite, note that $\mathcal{A}_2
    \subset \mathcal{A}_1$. From Bronez's theorem~\cite{BRO1988}, we have
    $\lambda_k(\mathcal{A}_2) < \lambda_k(\mathcal{A}_1)$, for $1 \le k \le
    \gls{numtaper}$, where $\lambda_k(\mathcal{A}_1)$ and
    $\lambda_k(\mathcal{A}_2)$ are the eigenvalues of the GPSS matrices for
    $\mathcal{A}_1$ and $\mathcal{A}_2$, respectively. Therefore,
\begin{align*}
    \lambda_k(\mathcal{A}_1) - \lambda_k(\mathcal{A}_2)
     & = \frac{\mathbf{x}_k^* \mathbf{R}(\mathcal{A}_1) \mathbf{x}_k}
    {\mathbf{x}_k^* \gls{gmatB} \mathbf{x}_k} -
    \frac{\mathbf{x}_k^* \mathbf{R}(\mathcal{A}_2) \mathbf{x}_k}
    {\mathbf{x}_k^* \gls{gmatB} \mathbf{x}_k}                         \\
     & = \frac{\mathbf{x}_k^* \mathbf{R}(\gls{subB}) \mathbf{x}_k}
    {\mathbf{x}_k^* \gls{gmatB} \mathbf{x}_k} > 0,
    \numberthis%})
\end{align*}
where the asterisk $*$ denotes complex conjugate transposition, and \gls{B}
is the entire signal band of any \gls{sig}. Since $\gls{gmatB}$ in the
denominator is positive definite, the numerator $\mathbf{x}_k^*
\mathbf{R}(\gls{subB}) \mathbf{x}_k$ must be positive. Thus,
$\mathbf{R}(\gls{subB})$ is a positive definite Hermitian matrix.

\section{Sidelobe Leakage in Missing Data Sampling}\label{app:sidelobe_leakage}
Missing data sampling presents an interesting special case within the
proposed framework. If the original sampling frequency is known, the optimal
taper sequences (\gls{gpss}) for analysis bands \gls{A0} and \gls{Ai},
denoted as \gls{wk0} and \gls{wki}, can be related by a simple frequency
shift operator
\gls{Ei}~\eqref{eq:freq_shift_op}~\cite{BRO1985,BRO1988,CHA2019}. Using the
identity described in Section~\ref{sec:subopt_measure}, the \gls{gep}
equation for the nominal band \gls{A0} given by~\eqref{eq:gep_at_A0} is
equivalent to
\begin{align*}
	\gls{Ei}^* \gls{gmatAi} \gls{Ei} \gls{wk0} =
	\gls{lambdak0} \gls{gmatB} \gls{wk0}.
	\numberthis
\end{align*}
Multiplying both sides by \gls{Ei} to the left and using the property
$\gls{Ei} \gls{Ei}^* = \gls{Ei}^* \gls{Ei} = \gls{idmat}$, we obtain
\begin{align*}
	\gls{gmatAi} \left[\gls{Ei} \gls{wk0}\right] =
	\gls{lambdak0} \left[\gls{Ei} \gls{gmatB} \gls{Ei}^*\right]
	\left[\gls{Ei} \gls{wk0}\right].
	\numberthis
\end{align*}
It can be shown~\cite{CHA2019} that for missing-data sampling with
normalized sampling frequency 1 Hz and $f_{\max} = 0.5$ Hz, the \gls{gpss}
matrix \gls{gmatB}~\eqref{eq:gpss_matrix_b} becomes the identical matrix
\gls{idmat}. Thus, $\gls{Ei} \gls{gmatB} \gls{Ei}^* = \gls{gmatB}$ and the
shifted tapers $\gls{wki} = \gls{Ei} \gls{wk0}$ satisfy the optimal criteria
for bias~\eqref{eq:bias_measure} and
variance~\eqref{eq:bound_factor_approx}. Moreover, since $\gls{lambdak0} =
\gls{lambdaki}$, sidelobe leakage \gls{E}~\eqref{eq:subopt_measure} is
minimized (i.e., $\gls{E} = 0$).

However, if the original sampling frequency is incorrectly estimated, the
sidelobe leakage is generally not optimized. For example, in the numerical
experiments, the actual sampling rate was 1.2 Hz, but the normalized
sampling frequency was assumed to be 1 Hz. This mismatch resulted in
non-zero sidelobe leakage across the signal band.  We formalize this
situation in the following theorem:

\begin{theorem}\label{thm:sidelobe_leakage_missing_data} For a missing data
	sampling scheme with an actual sampling frequency of $1/\gls{beta}$ Hz,
	where $\gls{beta} > 0$, normalized frequency of 1 Hz, and $\gls{fmax} =
	1/2$ Hz, the sidelobe leakage estimate $\gls{hatLki} = |\gls{lambdak0} -
	\gls{hatLki}|$ in~\eqref{eq:subopt_measure} for the shifted tapers
	$\hat{w}_k^i = \gls{Ei} \gls{wk0}$ is bounded lower by
	\begin{align*}
		\gls{hatLki} \ge 0
		\numberthis
		\label{eq:lower_bound_sidelobe_leakage}
	\end{align*}
	and upper by
	\begin{align*}
		\gls{hatLki} \le \gls{lambdak0}
		\left(1 +  \frac{f_w}{\Vert \gls{wk0} \Vert^2 - f_w}\right) < \infty,\, 
		\text{assuming}\, \gls{halfbw} < \| \gls{wk0} \|^2 \leq 2\gls{halfbw}.
		\numberthis
		\label{eq:upper_bound_sidelobe_leakage}
	\end{align*}
\end{theorem}
\begin{proof}
	To show the lower bound, we start with:
	\begin{align*}
		\hat{\lambda}_k^i & = \frac{{(\hat{w}_k^i)}^* \gls{gmatAi} \hat{w}_k^i}
		{{(\hat{w}_k^i)}^* \gls{gmatB} \hat{w}_k^i}
		= \frac{{\gls{wk0}}^* \gls{gmatA0} \gls{wk0}}
		{{(\gls{Ei} \gls{wk0})}^* \gls{gmatB} (\gls{Ei} \gls{wk0})}                                               \\
		                  & = \frac{\gls{lambdak0} \cdot 2f_w}
				  {{\gls{wk0}}^* \mathbf{ER}(\mathcal{B}) \gls{wk0}},
		\numberthis
		\label{eq:lambda_hat_ki}
	\end{align*}
	where $\mathbf{ER}(\mathcal{B}) \triangleq \gls{Ei}^* \gls{gmatB}
		\gls{Ei}$, whose elements under the missing data sampling
	scheme are given by
	\begin{align*}
		\mathbf{ER}(\mathcal{B} | n, m) & = \mathtt{e}^{-j 2 \pi \gls{fci} \gls{beta} k(n, m)}
		\gls{gmatBnm}
		\numberthis
		\label{eq:ERB_elements}
	\end{align*}
	and $k(n, n) = 0$, $|k(n, m)| \in \mathbb{Z}^+$ a positive integer for
	$n \neq m$, and $k(n,m) = -k(m,n)$, depending on a specific sampling
	scheme.  If $\gls{beta} \in \mathbb{Z}$ is an integer, it is
	clear~\cite{CHA2019} that $\mathbf{ER}(\mathcal{B}) = \gls{gmatB} =
	\gls{idmat}$ and thus $\hat{\lambda}_k^i = \gls{lambdak0}$, implying
	$\gls{hatLki} = 0$.

	To show the upper bound, let $P = {\gls{wk0}}^* \mathbf{ER}(\mathcal{B})
	\gls{wk0} = \| \gls{wk0} \|^2 + S_f$.  Here, $\| \gls{wk0} \|^2 =
	\sum_{n = 1}^N \gls{wk0}(n)^* \gls{wk0}$ is the sum of diagonal terms of
	$P$ and $S_f$ is the sum of off-diagonal terms given by
	\begin{align*}
		S_f &= \sum_{n \neq m} \mathbf{ER}(\mathcal{B}|n,m) 
		{\gls{wk0}}^*(n) \gls{wk0}(m).
		\numberthis
	\end{align*}
	Since $P^* = P$, $S_f$ is real-valued.  Suppose $S \triangleq
	\sum_{n \neq m} \gls{gmatBnm} {\gls{wk0}}^* \gls{wk0}$, then $S_f$ is
	the modulated version of $S$ with phase rotation.  From the
	normalization requirement~\eqref{eq:bg_norm_requirement}, we have
	$|S| = 2\gls{halfbw} - \|\gls{wk0}\|^2$, assuming $\|\gls{wk0}\|^2
	\leq 2\gls{halfbw}$.

	In the worst-case destructive interference scenario, where the
	modulation of the off-diagonal terms to align in phase opposition to the
	diagonal contribution, minimizing the total value of $P$, we have $S_f
	\ge -|S| = \|\gls{wk0}\|^2 - 2\gls{halfbw}$. Therefore, assuming
	$\gls{halfbw} < \|\gls{wk0}\|^2$ and using triangle inequality, we
	have the upper bound of $\gls{hatLki}$ given by
	\begin{align*}
		\gls{hatLki} & \leq |\gls{lambdak0}| + |\gls{hatLki}| 
		= \gls{lambdak0} \left( 1 + \frac{2\gls{halfbw}}
		{|P|}\right) \\
		& = \gls{lambdak0} \left( 1 + \frac{2\gls{halfbw}}
		{\left|\|\gls{wk0}\|^2 + S_f\right|}\right) \\
		& \leq \gls{lambdak0} \left( 1 + \frac{2\gls{halfbw}}
		{\left|\|\gls{wk0}\|^2 + \left( \|\gls{wk0}\|^2 - 2\gls{halfbw}\right)\right|}\right) \\
		& = \gls{lambdak0} \left( 1 + \frac{\gls{halfbw}}
		{\|\gls{wk0}\|^2 - \gls{halfbw}}\right) < \infty.
		\numberthis
		\label{eq:upper_bound_sidelobe_leakage_proof}
	\end{align*}
\end{proof}

This result shows that while frequency-shifted GPSS tapers can be optimal
under ideal conditions, mismatches in sampling frequency introduce bounded
but non-negligible sidelobe leakage.  The assumption $\gls{halfbw} <
\|\gls{wk0}\|^2 \leq 2\gls{halfbw}$ is typical since $\gls{gmatBnm} = 1$ for
$n = m$, and $\gls{gmatB}$ is positive definite.  Finally, we note that the
tapers \gls{wk0} are optimal \gls{gpss} for the nominal band \gls{A0}, but
may not be equivalent to \gls{mdss} used in~\cite{CHA2019}.

% **************
% references
% **************
\addcontentsline{toc}{References}{References}
\bibliographystyle{elsarticle-num}
\bibliography{fast_mtnufft}

\end{document}